# Facile Salt-Assisted Hydrothermal Synthesis of Nanodiamonds from CHO Precursors: Atomic-Scale Mechanistic Insights

Soumya Pratap Tripathy,[1] Sayan Saha,[1] Saurabh Kumar Gupta,[1] Pallavee Das,[1] Binay Priyadarsan Nayak,[1] Anup Routray,[1] Priya Choudhary,[1] Srihari V,[1] Bitop Maitra,[1] Ashna Reyaz,[1] Anushka Samant,[1] Debopriya Sinha,[1] Kritideepan Parida,[1] Kuna Das,[1] Abhijeet Sahoo,[1] Kunal Pal,[1] Sirsendu Sekhar Ray[1,2]*

[1]Department of Biotechnology and Medical Engineering, National Institute of Technology Rourkela, Rourkela, Odisha 769008, India
[2]Department of Biomedical Engineering, NEHU, Shillong, Meghalaya, 793022, India
*Corresponding author: raysirsendusekhar@gmail.com

**Abstract**

Hydrothermal synthesis offers an economical and scalable way to produce nanodiamonds under relatively mild, low-pressure and low-temperature conditions. However, its sustainability and the detailed mechanisms behind diamond formation in such environments are still not fully understood. In this work, we designed ten hydrothermal synthesis protocols using different CHO-based molecular precursors containing –COOH and –OH groups, such as organic acids, polyols, sugars, and polysaccharides. The reactions were carried out at 190°C in chlorinated, strongly alkaline aqueous solutions with alkali and alkaline-earth metal ions. Using high-resolution transmission electron microscopy (HRTEM) and X-ray photoelectron spectroscopy (XPS), we confirmed the presence of diamond-specific lattice planes and sp³-hybridized carbon structures. Our results show that the type of precursor, its molecular size, and the ionic composition of the solution play key roles in determining the defect patterns and polymorph distribution in the resulting nanodiamonds. Atomic-scale imaging showed both coherent and incoherent transitions from graphite to diamond, along with gradual lattice compression and complex twinning patterns. These observations provide direct insight into how interfacial crystallography and defect dynamics drive diamond formation in aqueous systems. Overall, the study positions hydrothermal synthesis as a sustainable, chemistry-driven, and tunable approach for creating nanodiamonds tailored for applications in quantum technologies, biomedicine, catalysis, and advanced materials.

## Introduction

Despite decades of effort, synthesizing diamonds under low-pressure and low-temperature (LPLT) conditions remains a major challenge, particularly when coupled with the need for controlled defect engineering [1–3]. Defects, however, are critical to the functionality of many diamonds, as they disrupt the $sp^3$ lattice symmetry and enable unique optical, electronic, mechanical, and catalytic properties [4,5]. For example, nitrogen–vacancy and silicon–vacancy centers are key to quantum sensing and single-photon emission; boron doping introduces p-type conductivity useful for electronic and electrochemical devices; and lonsdaleite or faulted diamond structures can even exceed conventional diamond in hardness[6–9]. Achieving controlled defect engineering under mild LPLT conditions would therefore open the door to sustainable, tunable routes for creating diamonds tailored for next-generation technologies—ranging from quantum systems and catalysts to biomedical devices, electronics, photonics, and biotechnology.

Although several techniques such as hydrothermal [10–12], laser [13], microwave [14], reduction–pyrolysis [15,16], electron beam [17], and plasma-based [18] methods have been explored for achieving diamond synthesis under LPLT conditions, the hydrothermal approach offers some clear advantages. It allows materials to form under much milder temperatures and pressures, offering better control over the chemical environment than high-energy techniques like laser, microwave, pyrolytic, or plasma processes, which often produce heterogenous carbon phases. In addition, because it is solution-based, scalable, and environmentally friendly, the hydrothermal method provides a safer and more sustainable pathway for material synthesis [19,20]. Previous studies, for example, have reported the formation of nanodiamonds smaller than 4 nm from nitrated graphene oxide or polyaromatic hydrocarbons in alkaline media (NaOH, 150 °C, 10 h), where nitro groups and sodium hydroxide ions promoted the transition from $sp^2$ to $sp^3$ bonding [10]. Similarly, other reports demonstrated nanodiamond formation from EDTA or glucose in acidic environments (HCl, 200–220 °C, 12–24 h), emphasizing the crucial role of precursor chemistry and ionic composition in the process [21,22].

Despite these encouraging results, the field remains largely underexplored. There has yet to be a systematic effort to validate the hydrothermal process as a reliable and sustainable route for nanodiamond synthesis or to fully explain how factors such as precursor type and ionic environment control defect formation and morphology. Moreover, the fundamental mechanisms governing diamond nucleation and growth under such mild LPLT conditions remain poorly understood—leaving significant gaps in our knowledge of how carbon transforms into diamond at the atomic scale.

In this study, we explored how different CHO-based precursors—such as carboxylic acids, polyols, and carbohydrates—interact with various alkali, alkaline-earth, and halide salts under mild hydrothermal conditions (190 °C) to better understand and validate the hydrothermal route for nanodiamond synthesis. Using high-resolution techniques like HRTEM and XPS, we confirmed the formation of nanodiamonds and identified their different polymorphs. Atomic-resolution imaging revealed both coherent and incoherent graphite-to-diamond interfaces, as well as gradual contraction of d-spacings, line defects, lattice distortions, and hierarchical twinning motifs, signifying the dynamic restructuring of carbon networks. Collectively, these observations suggest that CHO precursors first polymerize and carbonize into $sp^2$-rich intermediates, followed by localized $sp^2$ to $sp^3$ rehybridization driven by chemical precursors, ionic catalysis, and the unique thermodynamic conditions of the hydrothermal process.

## Results and discussion

The CHO-based precursors were grouped according to their dominant chemical functionalities. Group 1, rich in carboxylic acid (–COOH) groups, included malic acid (MA), citric acid (CA), and oxalic acid (OA). Group 2 consisted of low-molecular-weight polyols—glycerol (GLY) and ethylene glycol (EG). Group 3 contained hydroxyl-rich (–OH) precursors, represented by the monosaccharide dextrose (DX) and the glucose-derived polymers cellulose (CL) and starch (ST). Hydrothermal synthesis conducted under standardized conditions (190

°C, four h; Equations 1–10, Fig. 1) typically yielded whitish-gray precipitates. A distinct behavior was observed for Group 3 precursors. When reacted with NaOH and NaCl, they yielded black carbonaceous solids, indicative of amorphous or graphitic carbon formation. Interestingly, when these same precursors were pre-oxidized with $H_2O_2$, the products turned whitish or gray, resembling the outcomes seen for other CHO-based systems. The resulting precipitates of all the samples were dried, dispersed in isopropanol, and deposited on carbon-coated copper grids for HRTEM analysis.

At low magnification, samples displayed heterogeneous morphologies, including spherical, annular, spike-like, and broom-like features, as well as irregular agglomerates (Fig. S2A i–J i). Higher magnification revealed nanocrystalline domains (4–15 nm) embedded within amorphous matrices (Fig. S2A iii–J iii). FFT–IFFT analysis identified lattice spacings clustered around 2.18Å, 2.06Å, 1.93Å, 1.78Å, 1.50Å, 1.26Å, 1.17Å, and 1.07 Å (±0.03 Å), consistent with different twinned or faulted nanodiamond polymorphs[7,23]. The 2.06 Å spacing, corresponding to the {111} plane of cubic diamond, appeared ubiquitously across samples (Fig. 1, Table S1). Additional spacings larger than 3.0 Å, and in the 2.60–2.90 Å and 2.30–2.56 Å ranges, suggested the coexistence of graphitic layers, diamond–graphite interfaces, and stacking faults (Figs. S7–S26).

To further explore how reaction conditions influenced nanodiamond structure, thirty particles from each sample were examined for their d-spacing distribution and classified into the known nanodiamond polymorphs (Fig. 1A, Table 1) [7,23]. Particles with 1.78 Å spacings were categorized as n-type nanodiamonds, those with 2.18Å, 1.93Å, 1.50Å, or 1.17 Å spacings as hexagonal, and the remainder with 2.06 Å spacings as cubic [24]. Distinct group-dependent trends were observed. Group 1 (–COOH-rich) precursors generated low-to-moderate cubic fractions (17–33%) and moderate-to-high hexagonal fractions (27–43%), while $Ba^{2+}$ additives strongly promoted n-diamond formation (up to 70%). In Group 2 (polyols), GLY + NaOH + NaCl produced the highest cubic yield (43%), whereas EG combined with $Ca^{2+}$ or $Mg^{2+}$ favored n-diamond formation (67–70%) with a lower hexagonal fraction. For Group 3, the structural outcomes depended on precursor size and composition: starch promoted hexagonal phases (63%), cellulose favored n-diamond (70%), and dextrose enriched the cubic fraction (43%). Detection of larger spacings (>3.0 Å, 2.60–2.90 Å, and 2.30–2.56 Å) confirmed the presence of graphitic, onion-like [25], and faulted carbon structures. These secondary features varied systematically: NaCl with –OH–containing precursors or polymers boosted such carbon phases, whereas $Ba^{2+}$ suppressed them. Collectively, these observations emphasize the strong sensitivity of nanodiamond structure to hydrothermal chemistry, demonstrating the method's versatility in tailoring nanodiamond properties for specific applications.

Morphological analysis (Fig. 2B, Table S1) showed particle sizes ranging from 61.8 Å (MA + NaOH) to over 105 Å (EG + NaOH + CaCl$_2$ and CL + NaOH + NaCl). Eccentricity values (0.45–0.57; Fig. 2C) indicated near-spherical particles, though chloride ions tended to induce anisotropic growth. Smaller, more uniform particles were typically produced from simpler CHO precursors, while $Ba^{2+}$, $Mg^{2+}$, and polymeric systems yielded larger, more irregular particles (> 9 nm). Correlation analysis revealed that the cubic phase fraction decreased with increasing particle size (r = −0.69, p = 0.027), whereas the n-diamond fraction increased (r = 0.69, p = 0.028). Multiple regression ($R^2$ = 0.87) confirmed that the relative proportions of hexagonal and n-type phases were strong predictors of particle size. Interestingly, this trend differs from that observed in detonation-synthesized nanodiamonds, where n-type structures are more common in smaller particles[26]. This contrast suggests that the slower reaction kinetics of the hydrothermal process play a decisive role in shaping both morphology and defect evolution.

X-ray Photoelectron Spectroscopy (XPS) analysis revealed the elemental composition and bonding states of the hydrothermal precipitates. The products formed a carbon–salt matrix primarily containing carbon (40–77%), Na/Ca/Mg/Ba (8–17%), and oxygen (10–36%), with trace chlorine (<2%). Survey and high-resolution C 1s spectra (Fig. S1), along with element-specific scans (Figs. S7–S26), confirmed that $sp^3$-hybridized carbon (284.4–284.8 eV) dominated all samples, indicating the prevalence of the diamond phase. Its abundance ranged from 40.6% (MA + BaCO$_3$ + BaCl$_2$) to 70.1% (GLY + NaOH + NaCl). Minor $sp^2$-carbon contributions (≤ 5.5%) were observed in select systems (MA + NaOH, CA + OA + NaOH + NaCl, GLY + NaOH + NaCl, DX + NaOH + NaCl). The lower sp² signal compared with HRTEM observations reflects methodological differences, as XPS probes averaged surface chemistry whereas HRTEM resolves local crystalline domains. Oxygen-functionalized species were consistently detected, with C 1s peaks at ~288.0 eV (–COOH) and ~285.0 eV (C=O/O–C–O), supported by O 1s peaks at ~531.0 eV (C=O), ~532.5 eV (C–O), and ~533.5 eV (adsorbed $H_2O/OH$), consistent with oxidative surface termination. Trace nitrogen species (C–$NH_2$, C≡N, pyridinic N) appeared in most samples, likely introduced from minor impurities or atmospheric nitrogen. These oxygen- and nitrogen-functional groups are expected to modulate the surface energy, dispersion stability, and interfacial reactivity of the resulting nanodiamonds.

To evaluate the structural quality of the nanodiamonds and elucidate the transformation pathway from CHO precursors to diamond under hydrothermal conditions, ultra-high-resolution transmission electron microscopy (Ultra-HRTEM) was employed. For visual clarity, key lattice regions were color-highlighted, and diffuse FFT spots—representing the gradual contraction of graphitic spacings into distinct diamond lattice values—were displayed within the relevant ranges.

Hydrothermal synthesis using the EG + NaOH + $CaCl_2$ system provided direct atomic-scale evidence of the graphite-to-diamond (G-to-D) transformation, driven by lattice contraction and interfacial reorganization. Ultra-HRTEM images (Fig. 2D) and their corresponding FFT patterns (Fig. 2E) revealed lattice spacings of 3.2–3.5 Å, 2.02–2.2 Å, and 1.27–1.33 Å, which correspond to graphitic layers and the {111} and {2$\overline{2}$0} planes of diamond, respectively. IFFT mapping (Fig. 2B and 2F) further identified stacking-faulted regions where graphitic layers coherently transitioned into diamond-specific spacings (2.06 Å and 1.26 Å). This contraction behavior, consistent with previous observations in other diamond synthesis routes, is attributed to physicochemical stress under hydrothermal conditions [9,27,28]. Within these transition domains, stacked {111} and {2$\overline{2}$0} planes were observed to reorganize into a <001> face-centered cubic (FCC) lattice at interfacial seed boundaries (Fig. 2C). The conversion from sp²-bonded graphite to an sp³-hybridized diamond lattice is further substantiated by the <001> FCC diamond projection (Fig. 2A), which illustrates coherent connectivity between the {200}, {020}, {220}, and {2$\overline{2}$0} planes of diamond and the <0001> graphitic honeycomb lattice [29].

Similar atomic-scale features were observed in crystals synthesized from the EG + NaOH + $MgCl_2$ system. Ultra-HRTEM imaging (Fig. 2J), supported by FFT (Fig. 2H) and IFFT (Fig. 2K) analyses, confirmed a <01$\overline{1}$> cubic diamond orientation (Fig. 2L). Detailed FFT–IFFT inspection showed that the {0002} graphitic planes gradually contracted into {200} diamond planes (Fig. 2G), producing both coherent and incoherent G-to-D interfaces (Fig. 2G and 2I). Coherent domains indicated locally ordered lattice rearrangements during transformation, whereas incoherent regions exhibited atomic displacement and disorder, highlighting the kinetic sensitivity of the hydrothermal G-to-D conversion process [30]. Furthermore, wave-like lattice buckling observed within the diamond lattice (Fig. 2G) suggested a multi-axial stress-driven transformation toward lower d-spacings [31,32].

Fig. 3C presents an Ultra-HRTEM image of two fused nanodiamond crystals (D1 and D2) that nucleated independently from distinct graphitic domains (G1 and G2) in a post-hydrothermal sample synthesized using the CL + NaOH + NaCl system. A well-defined G-to-D boundary separates the two nucleation zones. FFT (Fig. 2H) and IFFT (Fig. 2E) analyses, along with Fig. 3D, 3G, and 3I, confirm that both D1 and D2 originated through incoherent G-to-D transitions from their respective graphitic precursors. As seen in earlier cases, the progressive contraction of d-spacings reflects the stress-induced transformation of graphitic layers into compact diamond lattices [25,33–35]. In domain G1, a bending-induced contraction was observed from 2.8–3.1 Å (magenta) to 2.7–2.8 Å (turquoise) and further to 2.3–2.6 Å (volt) (Fig. 3I). This transition produced a notch-like interfacial defect, which may facilitate or disrupt the nucleation of diamond-specific {111} planes in D1 (2.04–2.17 Å, blue). Similarly, D2 nucleated from stacked graphitic planes in G2, showing contraction from 2.7–3.18 Å (volt) and 2.7–3.17 Å (yellow) to 2.2–2.6 Å (turquoise), culminating in the formation of {111} diamond planes (2.04–2.17 Å, blue) (Fig. 3G). Notably, the {111} planes of D1 and D2 exhibit distinct orientations, fusing at an angle of ~96° between their respective {111} sets (Fig. 3F), suggesting a possible route for polycrystalline diamond growth during nanodiamond formation [36,37]. The region highlighted in Fig. 3B displays multilayered carbon phase distributions containing both graphitic and diamond d-spacings (Fig. 3Ai and 3Aii), indicative of graphitic–diamond heterostructures. The coexistence of these phases, consistent with earlier observations, underscores the role of heterogeneous nucleation and interfacial reorganization in driving diamond formation under hydrothermal conditions.

Fig. 4H shows a pentagonally twinned nanodiamond exhibiting both <01$\bar{1}$> rotation and reflection twins, identified in the post-hydrothermal precipitates obtained from the MA + NaOH + NaCl system. The corresponding crystallographic details are illustrated in the FFT (Fig. 4B) and IFFT images (Fig. 4A, 4E, and 4C). The structure comprises eight transient boundaries (TB1–TB8) separating seven distinct zones (Z1–Z7). Higher-magnification views of selected regions from Fig. 4E are presented in Fig. 4D, 4F, 4G, and 4I. Color-coded d-spacing maps in Fig. 4A and 4C reveal localized {111} diamond lattice nucleation and the emergence of twinned crystals from parallel graphitic G {0001} layers [38]. The graphite–diamond interface (zones Z1–Z4), highlighted in blue in Fig. 4A, coincides with twin boundaries TB3, TB5, and TB8 in Fig. 4H. Twin-related features are further evident: Fig. 4F shows TB3 and TB4 converging at a 71° junction in Z4, while Fig. 4G depicts TB7 arising from <01$\bar{1}$> rotational twinning, leading to a 71° rotation of {111} planes within the cubic diamond (CD) phase (Z6–Z7). Furthermore, line defects L1 and L2 along TB6 represent delocalized electronic states resembling graphene zigzag edges (Fig. 4D). Graphitic inclusions embedded within the diamond lattice (Fig. 4E, dashed box) suggest that graphite–diamond heterostructures naturally form during hydrothermal synthesis. At TB1, AA-stacked graphite transforms into AA-stacked hexagonal diamond (HD), which further converts to AB′-stacked CD in zone Z1 (Fig. 4I) [39–42]. In zone Z2, a similar AA-to-AB′ transition occurs, accompanied by wave-like lattice buckling (lime green ellipses, Fig. 4I). Such disordered heterophases, as previously reported, likely facilitate diamond nucleation and growth [43].

Fig. 5H shows an ultra-high-resolution TEM image of a nanodiamond crystal obtained from the MA + BaCO₃ + BaCl₂ system. The crystal displays four hierarchically stacked <01$\bar{1}$> reflection twins (z1–z4) separated by three twin boundaries (TB1–TB3), as confirmed by the FFT (Fig. 5B) and IFFT (Fig. 5E) analyses. The spatial lattice map (Fig. 5C) reveals distinct d-spacings of 4–6 Å (red), 2.5–3.5 Å (green), and 2.02–2.2 Å (blue), corresponding to planar carbon layers, compressed graphitic planes, and diamond {111} planes, respectively. The twinned crystal appears to have nucleated through stress-induced compression of stacked planar carbon layers into diamond {111} lattices via an intermediate graphitic phase (Fig. 5A and 5C). Residual graphitic inclusions observed in zones z1–z3 (Fig. 5D–5G) further support this pathway. The twin boundaries likely serve a stabilizing role by redistributing local stress and reducing the nucleation energy during crystal growth. In zone z4, the atomic lattice undergoes a coherent transformation from a <01$\bar{1}$> diamond lattice near TB3 to a <0001> graphitic honeycomb lattice (cyan), and back to a <01$\bar{1}$> diamond lattice (Fig. 5I). This results in a sandwiched configuration of graphene-like layers enclosed between cubic diamond regions. Such hybrid interfaces likely arise from incomplete phase transitions or residual internal stress during the hydrothermal process.

Therefore, atomic-scale HRTEM analysis revealed faulted graphite–diamond regions with progressive interlayer d-spacing contraction across different hydrothermal chemistries. Both coherent and incoherent G to D interfaces were observed, acting as nucleation sites for diamond domains and cubic facets. Localized physicochemical stresses, manifesting as lattice buckling, notches, and multi-axial distortions, compress graphitic planes into diamond lattices, promoting ordered lattice formation at interfaces [25,44]. Complex twin architectures, residual graphitic inclusions, and edge-state defects were also identified, which appear to play a dual role: enabling diamond nucleation while simultaneously accommodating strain and limiting crystal growth. Collectively, these observations suggest that chemically induced stresses, faults, and interfacial defects cooperatively drive the heterogeneous G to D phase transition during low-temperature hydrothermal nanodiamond formation.

The results indicate that nanodiamond formation under hydrothermal conditions probably proceeds through a two-step transformation: CHO-based precursors first evolve into sp² or graphitic domains, which subsequently convert into sp³ or nanodiamonds. During hydrothermal carbonization, CHO precursors undergo decarboxylation and dehydration to form reactive carbonaceous radicals. These intermediates condense into aliphatic, cyclic, and polycyclic frameworks, serving as the foundation for higher-order carbon structures[45–47]. The presence of alkali and alkaline-earth metals such as Na, Ca, and Mg further accelerates this process by catalyzing decomposition, enhancing cyclic/polycyclic condensation, and promoting dehydration-driven hydrophobic assembly into polymerized sp² carbons resembling graphene and graphite[48–51]. Once sp² domains are established, subsequent hydrogenation, hydroxylation, or chlorination disrupts the delocalized π-electron network, reducing the thermodynamic barrier for the sp²-to-sp³ transformation [52–54]. In a confined hydrothermal environment, reactive intermediates and localized stresses arising from heterogeneous carbon–salt matrices likely act synergistically to reorganize stacking sequences and promote interlayer bonding, ultimately driving the nucleation and growth of nanodiamonds [33,44,55,56]. Thus, hydrothermal nanodiamond synthesis emerges from a complex interplay among precursor structure, ionic chemistry, and phase-transition dynamics.

Importantly, this study demonstrates that a broad range of –COOH- and –OH-functionalized CHO precursors, from simple organic acids to complex biopolymers such as starch and cellulose, can be transformed into nanodiamonds under mild hydrothermal conditions. This marks a significant step forward in diamond synthesis, revealing that variations in precursor chemistry, functional groups, molecular architecture, and associated salts act as intrinsic chemical levers to modulate defect structures, polymorph distributions, and surface terminations. Despite the success of achieving diamond formation at low pressure and temperature, the resulting crystals still exhibit multiple lattice faults, similar to those observed in detonation-synthesized diamonds, indicating that the hydrothermal route, while effective, also preserves structural imperfections arising from the interplay of nucleation, stress, and defect dynamics. Further studies with larger datasets and controlled replicates are needed to elucidate additive–phase relationships and statistically guide predictive synthesis.

Optimizing the chemical environment and reaction parameters could help regulate these imperfections, enabling the synthesis of more uniform and application-specific nanodiamonds. Overall, the results provide evidence that nanodiamonds and their polymorphs can form at 190 °C under aqueous, low-pressure conditions, overturning the long-standing view that they are exclusive products of high-pressure, high-temperature events. [57–60] This advancement not only supports greener nanomanufacturing and biocompatible material design but also establishes hydrothermal synthesis as a sustainable and tunable platform for producing high-quality, application-oriented nanodiamonds.

**Conclusion**

This study establishes hydrothermal synthesis as a viable and energy-efficient alternative to conventional HPHT and detonation methods for synthesizing nanodiamonds and their polymorphs. The role of COOH/OH-functionalized CHO precursors and ionic catalysts, particularly alkali and alkaline earth metal salts and chlorides, in mediating phase transitions, nanoscale structural evolution, and morphology underscores the tunability of this synthesis approach. By elucidating the sequential transformation of organic precursors into graphitic intermediates and subsequently into sp³-hybridized nanodiamond structures, we provide atomic-scale mechanistic insights into chemically induced nanodiamond nucleation under low-pressure and low-temperature conditions. Advancements in hydrothermal nanodiamond synthesis could unlock process-controllable, scalable, cost-effective, and environmentally sustainable pathways to produce nanodiamonds. With potential applications spanning quantum sensing, drug delivery, and high-performance composites, continued research on reaction kinetics, precursor engineering, and post-synthesis functionalization will be vital to fully realize the promise of hydrothermally synthesized nanodiamonds.

## Methods

**Hydrothermal synthesis:** The samples were prepared by mixing the CHO-precursors with salts in distilled water. The aqueous mixture was kept sealed in a 50 mL Teflon-lined container, which was held within a mechanically restrained stainless steel hydrothermal reactor. The reactor was heated in a muffle furnace at 190 °C for 4 hours, followed by natural cooling to room temperature in ambient conditions. The post-hydrothermal samples are clear and transparent solutions with white/greyish precipitate. The precipitates were dried and characterized using HRTEM and XPS.

**MA+NaOH:** 4 g of malic acid (MA) was mixed with 14 g of NaOH in 35 ml of water for hydrothermal treatment.

**MA+NaOH+NaCl:** 4 g of malic acid (MA) was mixed with 8 g of NaOH and 6 g of NaCl in 35 ml of water for hydrothermal treatment.

**CA+OA+NaOH+NaCl:** 2 g of citric acid (CA) and 2 g of oxalic acid (OA) with 6 g NaCl and 8 g NaOH in 25 ml of water for hydrothermal treatment.

**MA+BaCO3+BaCl2:** 4 g of malic acid (MA) was mixed with 2 g of $BaCO_3$ and 2 g of $BaCl_2$ in 35 ml of water for hydrothermal treatment.

**GLY+NaOH+NaCl:** 10 mL of glycerol (GLY) was mixed with 11 g of NaOH and 3 g of NaCl in 20 ml of water for hydrothermal treatment.

**EG+NaOH+MgCl2:** 10 mL of ethylene glycol (EG) with 4 g of NaOH and 6g of MgCl2 in 20 mL of water for hydrothermal treatment.

**EG+NaOH+CaCl2:** 10 ml of ethylene glycol (EG) with 4 g of NaOH and 6g of $CaCl_2$ in 20 ml of water for hydrothermal treatment.

**DX+NaOH+NaCl:** Dextrose (DX) was oxidized before hydrothermal treatment using hydrogen peroxide ($H_2O_2$). 4 g of dextrose in 20 ml of water was mixed with 20 ml of $H_2O_2$ overnight to convert the dextrose to glucuronic acid and other smaller organic molecules. 8 g of NaOH and 6 g of NaCl were added to the oxidized reaction mixture and used for hydrothermal treatment.

**ST+NaOH+NaCl:** Starch (ST) was oxidized before hydrothermal treatment using hydrogen peroxide ($H_2O_2$). 4 g of starch in 20 ml water was mixed with 20 ml of $H_2O_2$ at 80 $^0$C under prolonged agitation through stirring with a magnetic stirrer for 12 hours for converting starch into hydrolyzed, polyhydroxy, and carboxylic acids containing smaller molecules. 8 g of NaOH and 6 g of NaCl were added to the oxidized reaction mixture and used for hydrothermal treatment.

**CL+NaOH+NaCl:** Cellulose (CL) was oxidized before hydrothermal treatment using hydrogen peroxide ($H_2O_2$). 4 g of cellulose in 20 ml of water was mixed with 20 ml of $H_2O_2$ at 80 $^0$C under prolonged agitation through stirring with a magnetic stirrer for 12 hours for converting starch into hydrolyzed, polyhydroxy, and carboxylic acids containing smaller molecules. 8 g of NaOH and 6 g of NaCl were added to the oxidized reaction mixture and used for hydrothermal treatment.

## Characterization

The dried precipitate was suspended in isopropanol and drop-casted onto a carbon-coated copper grid under vacuum. The grid was examined under a FEI Tecnai G2 TF30 S-Twin transmission electron microscope at 300 kV for high-resolution imaging. The HRTEM images were analyzed using ImageJ's standard FFT and IFFT process. The powdered sample's XPS spectrum was obtained using standard monochromatic Al K$\alpha$ X-ray radiation (1486 eV). The XPS data were plotted and analyzed using OriginLab's software.

## Statistics and Reproducibility

The error bars represent the standard deviation obtained from the population of independent experiments and/or measurements as stated in the manuscript. Pearson correlation and multiple regression analyses were used to investigate the relationship between the mean size or eccentricity of nanodiamonds and the distribution of their polymorphs.

## Data availability statement

All data needed to evaluate the conclusions are presented in the paper.

## Acknowledgments

The authors thank the National Institute of Technology Rourkela (NITR) for providing the laboratory and characterization facility. We also acknowledge support from Mr. Suman Sarkar and Mr. Subhabrata Chakraborty for the use of the HRTEM characterization facilities available at NITR. Authors also acknowledge the valuable contributions of the Indian Institute of Technology Roorkee and the Indian Institute of Technology Jammu for the XPS characterization facilities.

## Competing interests

Authors have no competing interests.

## Authors' contributions

**S.P.T.:** Investigation, Methodology, Investigation, Data curation, Formal analysis, Writing – Original Draft, Communication.

**S.S., P.D., and B.P.N.:** Investigation, Methodology, Sample preparation.

**S.K.G.:** Investigation, Data curation, XPS characterization, Writing – Review & Editing, Communication

**P.C., S.V., B.M., A.R., A.S., and D.S.:** Investigation, Methodology, Sample preparation (protocol refinement).

**A.R., K.P., K.D., A.S.:** Validation, Investigation, Data curation

**K.P.:** Investigation, Writing

**S.S.R.:** Conceptualization, Methodology, Investigation, Data curation, Formal analysis, Writing – Original Draft, Supervision, Project administration, Communication.

| Groups | Samples | c-ND * | h-ND ** | n-ND *** | n-ND*** and h-ND** |
|---|---|---|---|---|---|
| Group 1 | MA+ NaOH | 17% | 40% | 67% | 23% |
| | MA + NaOH+ NaCl | 33% | 27% | 30% | 10% |
| | CA+ OA+ NaOH +NaCl | 27% | 43% | 43% | 23% |
| | MA+BaCO$_3$+BaCl$_2$ | 17% | 40% | 70% | 27% |
| Group 2 | GLY+ NaOH+ NaCl | 43% | 20% | 30% | 10% |
| | EG+NaOH+CaCl$_2$ | 26% | 17% | 67% | 10% |
| | EG+NaOH+MgCl$_2$ | 26% | 17% | 70% | 13% |
| Group 3 | DX+ NaOH+ NaCl | 43% | 10% | 50% | 3% |
| | ST+NaOH+NaCl | 20% | 63% | 40% | 23% |
| | CL+ NaOH+ NaCl | 13% | 37% | 70% | 20% |

*Only 2.06Å (±0.03); excluding hexagonal(h) and new (n) nanodiamonds (ND); c means cubic

** If any of 2.18Å (±0.03),1.93Å (±0.03),1.5Å (±0.03)

*** If 1.78Å(±0.03)

**Table 1:** Nanodiamond polymorph distribution in post-hydrothermal precipitates from HRTEM lattice-spacing analysis of thirty randomly selected crystallites per sample

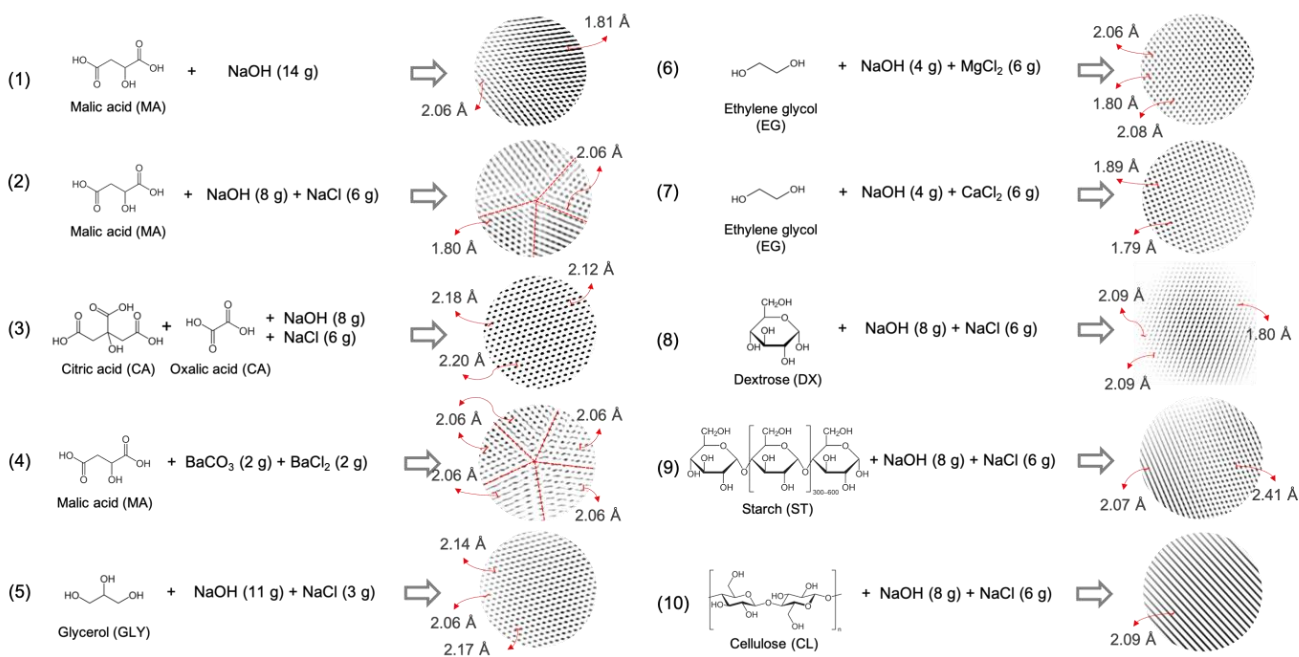

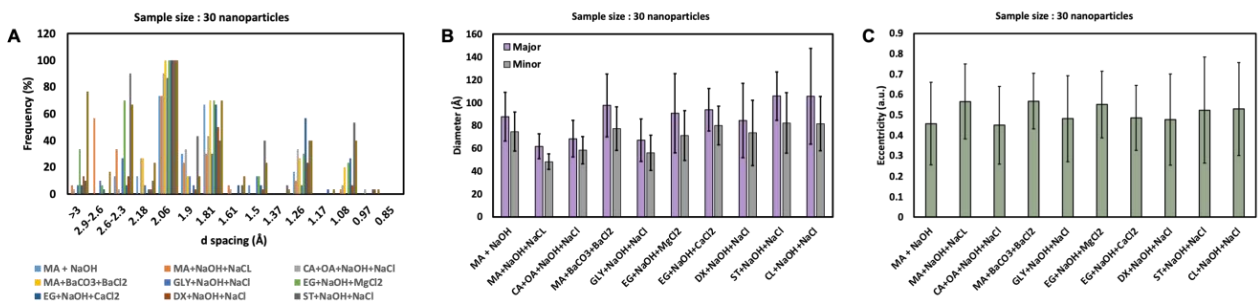

**Figure 1: Hydrothermal protocols and size, geometry, and d-spacings statistics from HRTEM Observations.** (1-10) Hydrothermal protocols followed for the synthesis of nanodiamonds. Snapshots of atomic lattices on the product side are extracted from HRTEM observation of nanoparticles in the post-hydrothermal samples. (A) d-spacings found in nanoparticles (n=30) found in HRTEM observation of post-hydrothermal samples. (B) Statistics of major and minor diameters of nanoparticles (n=30) found in HRTEM observation of post-hydrothermal samples. (C) Eccentricity of nanoparticles (n=30) found in HRTEM observation of post-hydrothermal samples.

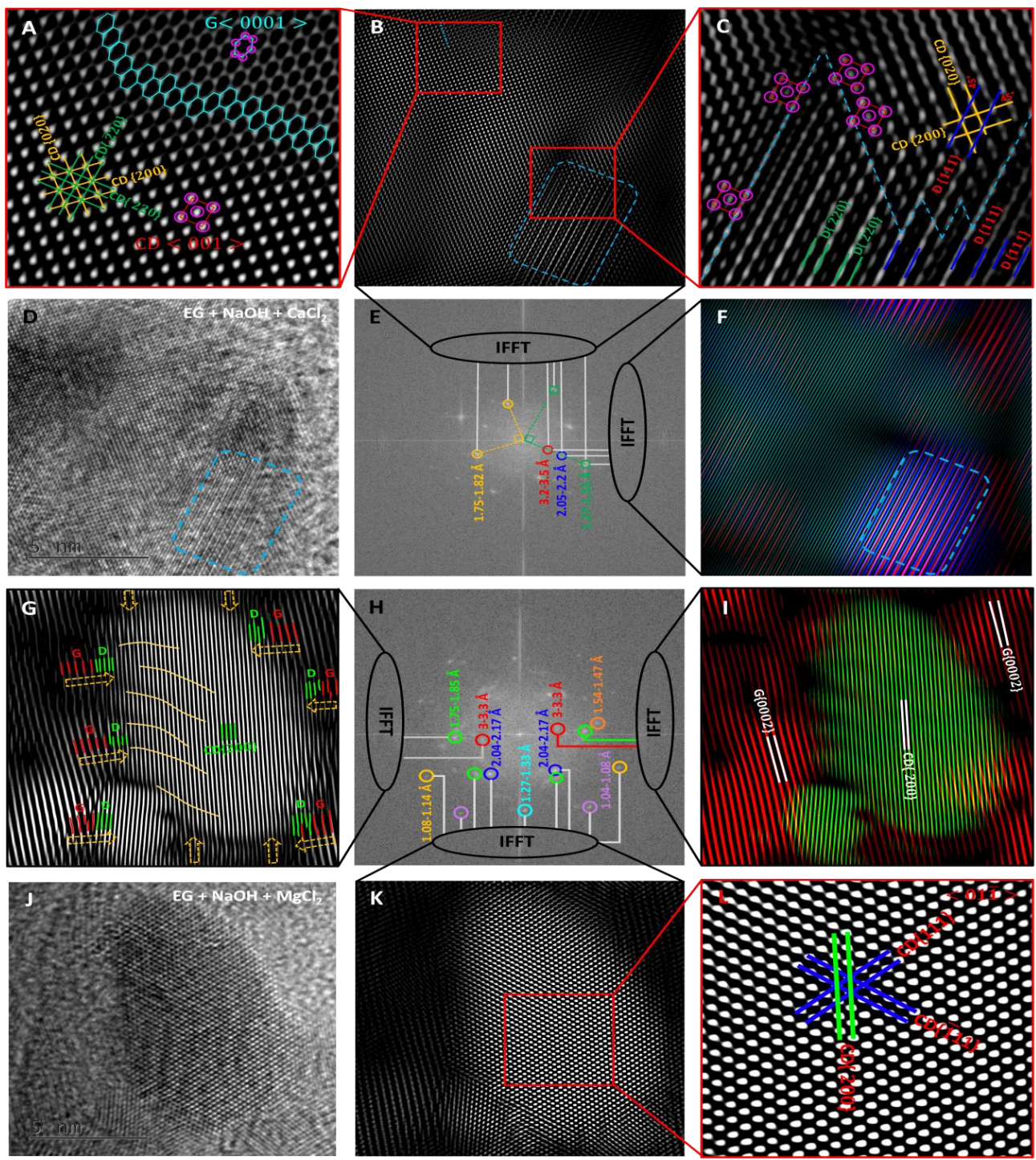

**Figure 2. Nanostructure and nucleation of nanodiamond crystals from EG+NaOH+CaCl₂ and EG+NaOH+MgCl₂ hydrothermal chemistries.**
(A) Face-centered cubic lattice (<001>) of the cubic diamond phase sharing a G-to-D boundary with graphitic (<0001>) domains. (B) Gray-scale IFFT of selected bright pixels from FFT (E), corresponding to d-spacings of 3.2–3.3 Å (red), 2.05–2.2 Å (blue), 1.27–1.33 Å (green), and 1.75–1.82 Å (golden yellow), with nucleation sites highlighted (dashed blue box). (C) Region under dashed line showing stacked graphitic and diamond phases with 2.05–2.2 Å (blue) and 1.27–1.33 Å (green) lattices. (D) Ultra-HRTEM of nanodiamond from EG+NaOH+CaCl₂; (E) corresponding FFT; (F) IFFT highlighting spatial distribution of 3.2–3.3 Å, 2.05–2.2 Å, and 1.27–1.33 Å lattices. (G–I) Images from EG+NaOH+MgCl₂ showing gradual lattice contraction in graphitic domains and embedded diamond phase. (J) Ultra-HRTEM of nanodiamond; (K) IFFT displaying multiple d-spacings (3.2–3.3, 2.04–2.17, 1.75–1.85, 1.08–1.14, 1.54–1.47, and 1.04–1.08 Å); (L) snapshot of cubic diamond <0$\bar{1}$1> lattice.

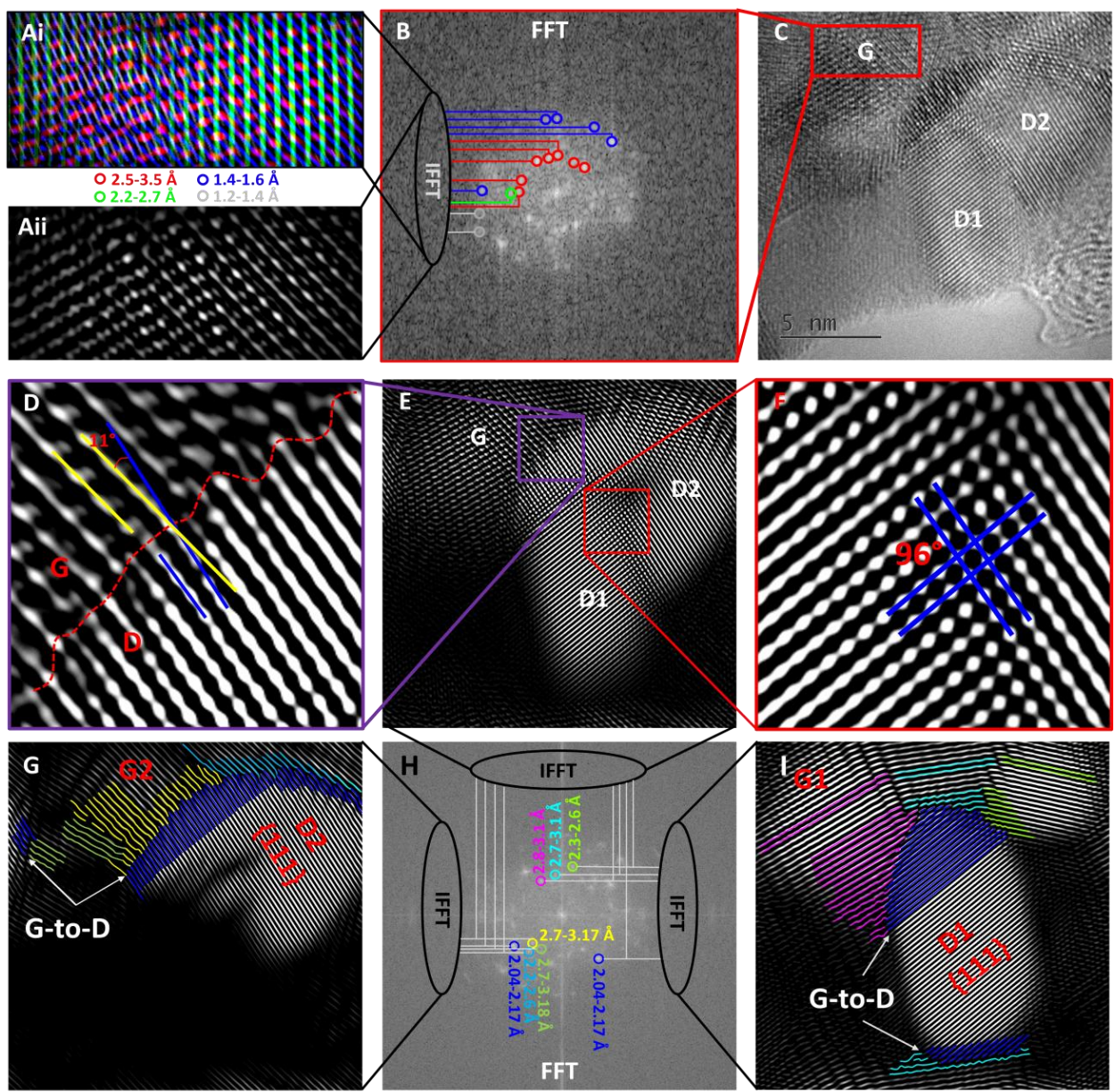

**Figure 3. Nanostructure and nucleation of two fused nanotwinned crystals from CL+NaOH+NaCl hydrothermal chemistry.**
(Ai) Color-coded IFFT image of selected bright pixels from FFT (B), corresponding to d-spacings of 2.5–3.5 Å (red), 2.2–2.7 Å (green), 1.4–1.6 Å (blue), and 1.2–1.4 Å (gray), showing spatial distribution of atomic lattices. (Aii) Gray-scale IFFT of the same regions. (B) FFT image of graphitic region (G) in (C). (C) Ultra-HRTEM of two fused nanotwinned nanodiamond crystals (D1, D2) nucleating from the graphitic phase (G). (D) Snapshot of incoherent G-

to-D2 interface. (E) Gray-scale IFFT image of selected bright pixels from FFT (H), corresponding to 2.7–3.17 Å (yellow), 2.8–3.1 Å (magenta), 2.7–3.1 Å (cyan), 2.37–3.18 Å (lime green), 2.3–2.6 Å (green), 2.2–2.6 Å (bright blue), and 2.04–2.17 Å (blue). (F) Fusion of D1{111} and D2{111} at ~96°. (G) Lattice transition from G2 to D2{111} via incoherent boundary. (H) FFT of (C). (I) Transition from G1{111} to D1 through incoherent G-to-D boundary with mismatch.

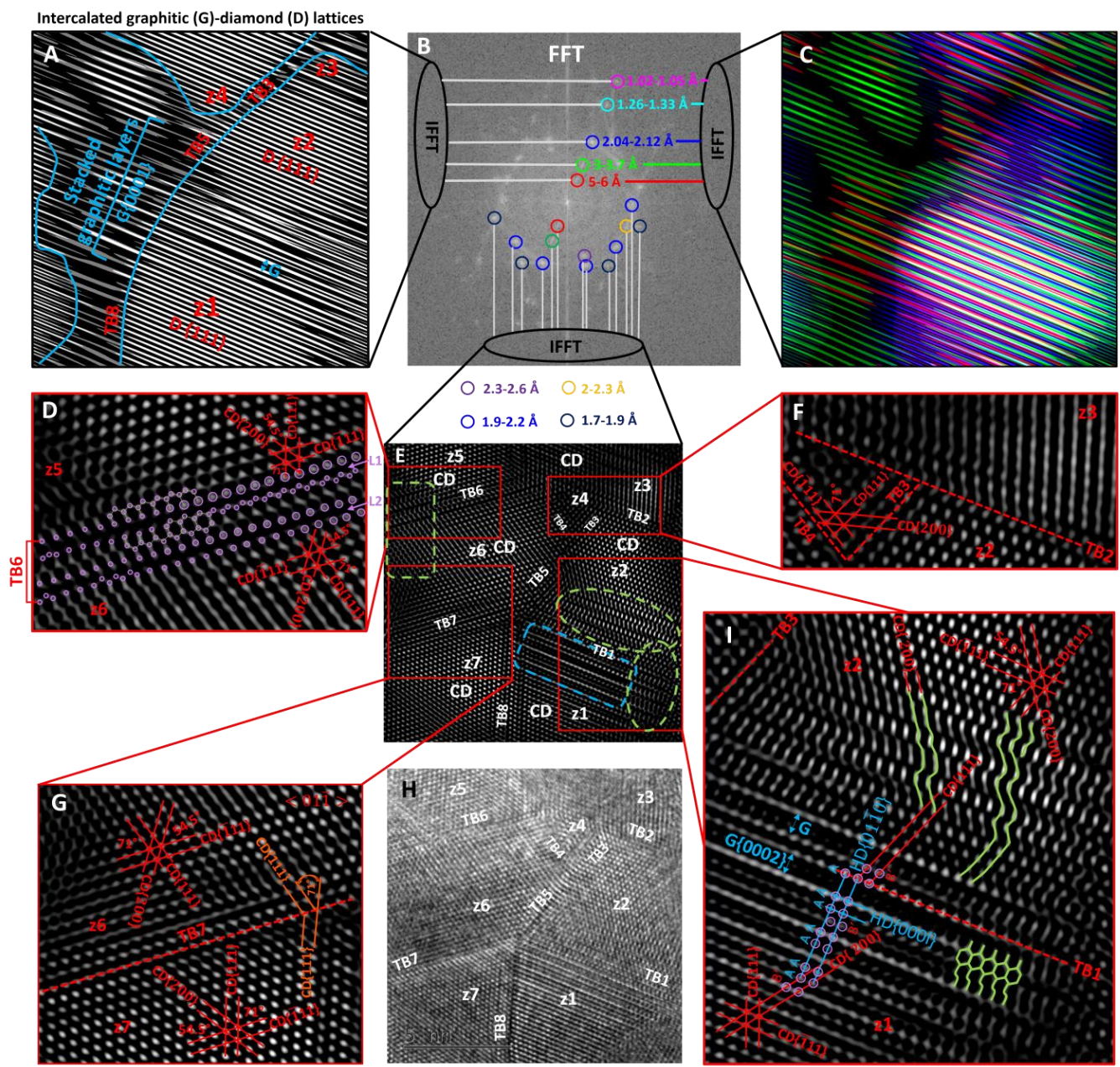

**Figure 4. Nanostructure and nucleation of a pentagonal-twinned nanodiamond crystal from the MA+NaOH+NaCl protocol.**
(A) IFFT image showing stacked interleaved graphitic and diamond lattices derived from bright pixels at 5–6 Å, 3–3.7 Å, 2.04–2.12 Å, 1.26–1.33 Å, and 1.02–1.05 Å on FFT (B). (B) FFT of the ultra-HRTEM image in (H). (C) Color-coded IFFT image of interleaved lattices: 5–6 Å (red), 3–3.7 Å (green), 2.04–2.12 Å (blue), 1.26–1.33 Å (turquoise), and 1.02–1.05 Å (magenta). (D) Atomic lattice of $z5$–$z6$ across twin boundary TB6, showing line defects formed by graphitized, partially graphitized, and diamond-specific phases. (E) IFFT of bright pixels at 2.3–2.6 Å (magenta), 2–2.3 Å (blue), 1.9–2.2 Å (blue), and 1.7–1.9 Å (black). Hexagonal diamond with graphitic defects (blue box) and disordered diamond (green ellipses) are highlighted. (F–G) Atomic lattices across twin boundaries TB2 ($z2$–$z3$) and TB7 ($z6$–$z7$). (H) Ultra-HRTEM of twinned nanodiamond. (I) Transition from cubic diamond (CD) to hexagonal diamond (HD) with interleaved graphitic phase in $z1$–$z2$, showing stacking shifts (AB′→AA/AB′ defects).

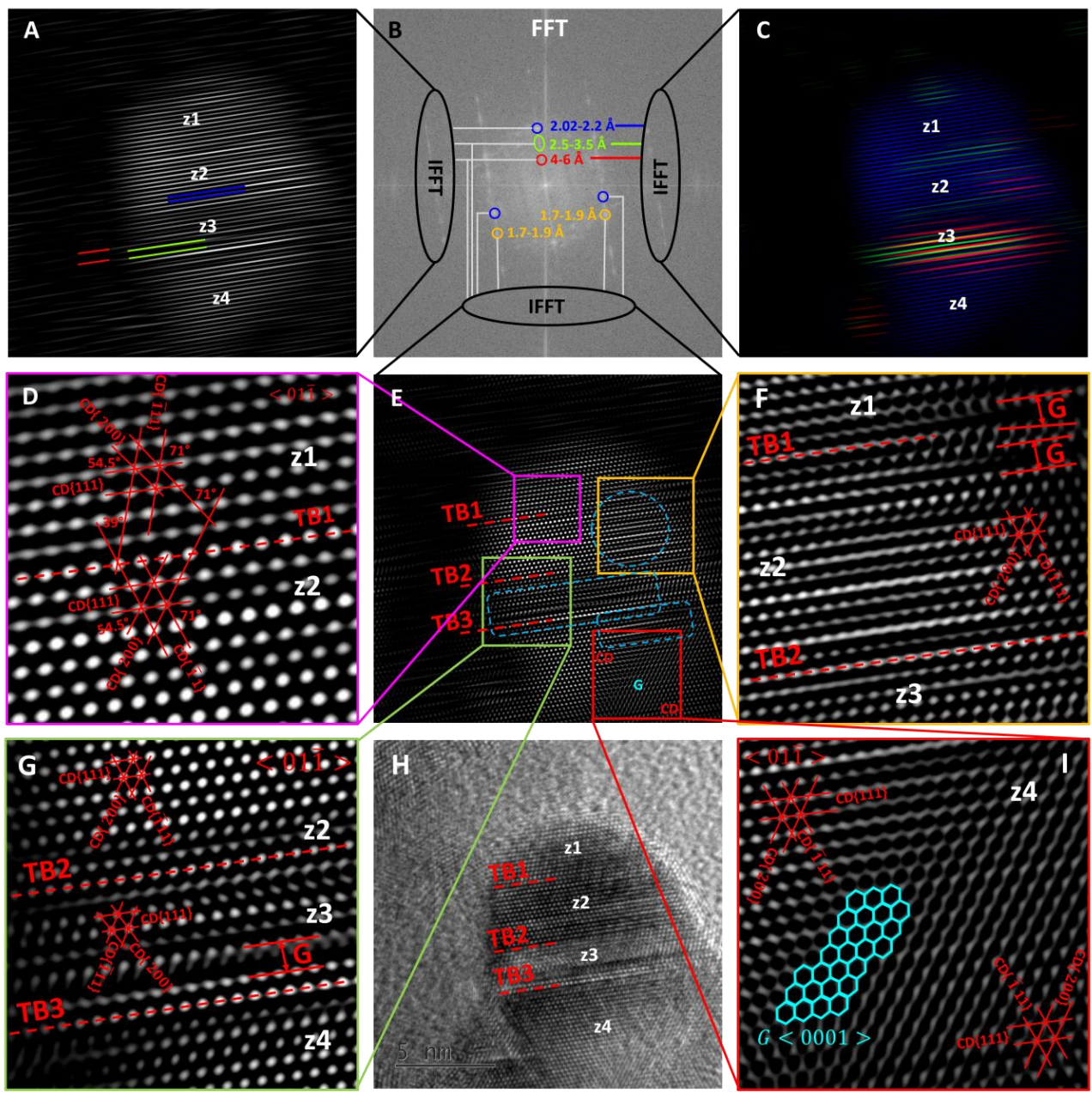

**Figure 5. Nanostructure and nucleation of a nanodiamond crystal with hierarchically stacked twins from MA+BaCO₃+BaCl₂ hydrothermal chemistry.**
(A) Gray-scale IFFT image of selected bright pixels from FFT (B), highlighting stacked interleaved graphitic and diamond lattices at 4–6 Å (red), 2.5–3.5 Å (green), and 2.02–2.2 Å (blue). (B) FFT of the ultra-HRTEM image in (H). (C) Color-coded IFFT showing spatial distribution of graphitic and diamond phases. (D) Atomic lattice of cubic diamond in zones z1 and z2 across the twin boundary TB1. (E) IFFT of bright pixels at 4–6 Å (red), 2.5–3.5 Å (green), 2.02–2.2 Å (blue), and 1.7–1.9 Å (golden yellow), with disordered stacked graphitic/diamond regions highlighted (blue rectangle and ellipses). (F–G) Atomic lattices of zones z1–z3 showing stacked graphitic and diamond phases, with TB2 and TB3 marking interfaces with adjacent zones. (H) Ultra-HRTEM of the nanodiamond crystal. (I) Zone 4 lattice containing a graphene-like phase (<0001>, cyan honeycomb) sandwiched between cubic-diamond domains (red lines).

# Supplementary Information

# Facile Salt-Assisted Hydrothermal Synthesis of Nanodiamonds from CHO Precursors: Atomic-Scale Mechanistic Insights

Soumya Pratap Tripathy,[1] Sayan Saha,[1] Saurabh Kumar Gupta,[1] Pallavee Das,[1] Binay Priyadarsan Nayak,[1] Anup Routray,[1] Priya Choudhary,[1] Srihari V,[1] Bitop Maitra,[1] Ashna Reyaz,[1] Anushka Samant,[1] Debopriya Sinha,[1] Kritideepan Parida,[1] Kuna Das,[1] Abhijeet Sahoo,[1] Kunal Pal,[1] Sirsendu Sekhar Ray[1,2*]

[1]Department of Biotechnology and Medical Engineering, National Institute of Technology Rourkela, Rourkela, Odisha 769008, India
[2]Department of Biomedical Engineering, NEHU, Shillong, Meghalaya, 793022, India
*Corresponding author: raysirsendusekhar@gmail.com

Table S1: Frequency distribution of lattice spacings in post-hydrothermal precipitates from HRTEM analysis of thirty randomly selected crystallites per sample

| d -spacings (Å) ±0.03 | Group 1 | | | | Group 2 | | | Group 3 | | |
|---|---|---|---|---|---|---|---|---|---|---|
| | MA+ NaOH | MA+ NaOH+ NaCl | CA+ OA+ NaOH +NaCl | MA+ BaCO3+ BaCl2 | GLY+ NaOH+ NaCl | EG+ NaOH + CaCl2 | EG+ NaOH + MgCl2 | DX+ NaOH +NaCl | ST+ NaOH +NaCl | CL+ NaOH +NaCl |
| Above 3 | 0 | 2 | 1 | 0 | 2 | 2 | 10 | 4 | 3 | 23 |
| 2.60-2.90 | 0 | 17 | 0 | 0 | 3 | 1 | 2 | 0 | 0 | 5 |
| 2.30-2.56 | 4 | 10 | 1 | 0 | 8 | 2 | 21 | 4 | 27 | 20 |
| 2.18 | 4 | 0 | 8 | 8 | 2 | 1 | 0 | 1 | 3 | 7 |
| 2.06 | 22 | 22 | 27 | 30 | 26 | 30 | 30 | 30 | 30 | 30 |
| 1.93 | 9 | 7 | 10 | 4 | 4 | 2 | 0 | 1 | 13 | 4 |
| 1.78 | 20 | 9 | 13 | 21 | 9 | 20 | 21 | 15 | 12 | 21 |
| 1.61 | 0 | 2 | 1 | 0 | 0 | 2 | 0 | 0 | 2 | 4 |
| 1.5 | 2 | 2 | 0 | 0 | 4 | 2 | 4 | 1 | 12 | 7 |
| 1.37 | 0 | 0 | 0 | 0 | 0 | 0 | 0 | 0 | 2 | 1 |
| 1.26 | 5 | 3 | 10 | 8 | 2 | 17 | 9 | 7 | 12 | 12 |
| 1.17 | 0 | 0 | 0 | 0 | 0 | 1 | 0 | 0 | 0 | 1 |
| 1.07 | 0 | 1 | 2 | 6 | 0 | 8 | 7 | 2 | 16 | 12 |
| 0.97 | 0 | 0 | 1 | 0 | 0 | 1 | 0 | 1 | 0 | 1 |
| Mean Size (Å) | 87.74± 21.23 | 61.81± 10.95 | 68.53± 16.10 | 97.6± 27.56 | 67.17± 18.59 | 93.8± 18.75 | 90.8± 34.74 | 84.34± 32.53 | 105.8± 21.09 | 105.63± 41.99 |
| Eccent-ricity | 0.46± 0.2 | 0.57± 0.18 | 0.45± 0.19 | 0.57± 0.14 | 0.48± 0.21 | 0.49± 0.16 | 0.55± 0.16 | 0.48± 0.22 | 0.52± 0.26 | 0.53± 0.21 |

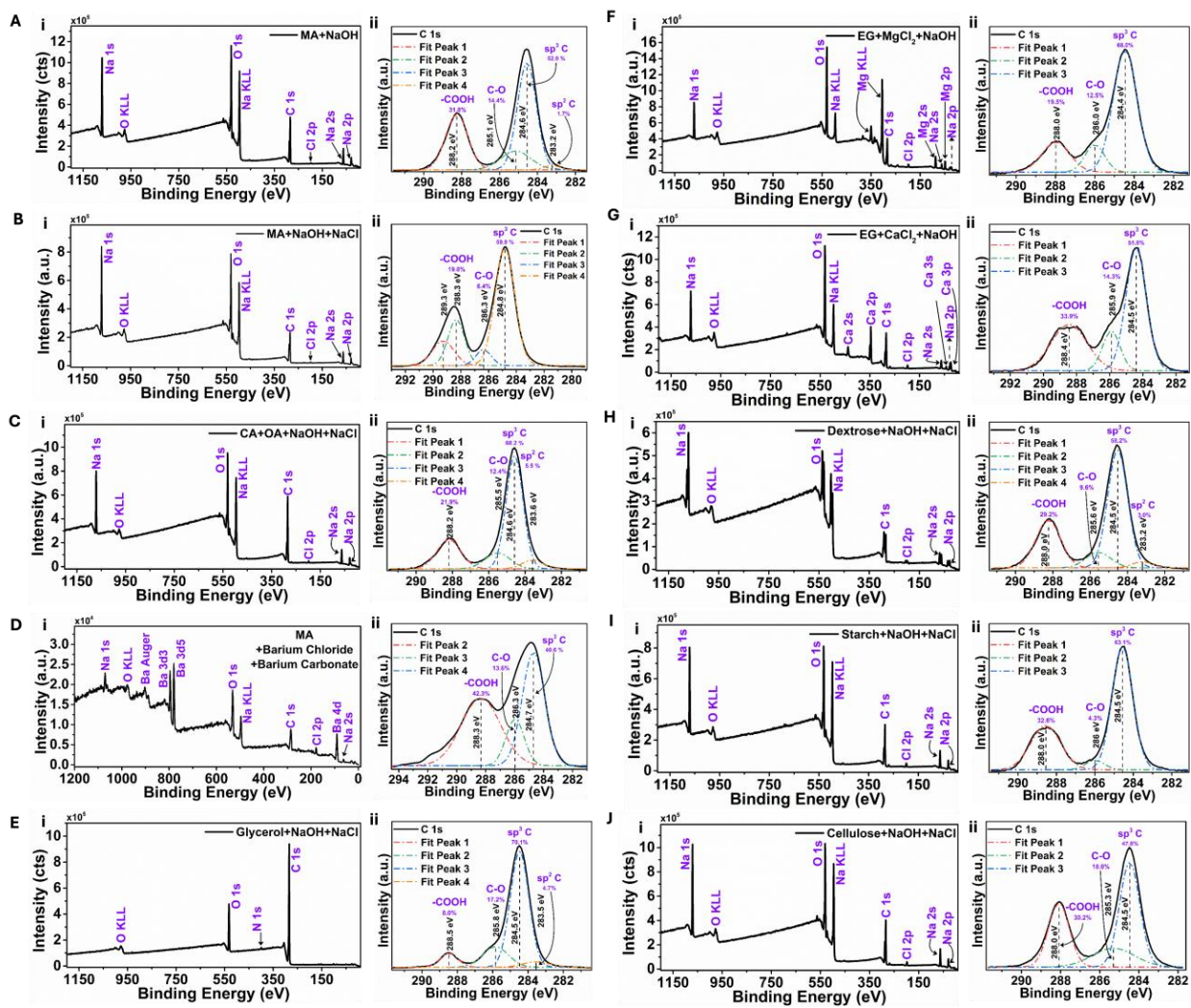

**Figure S1: XPS survey scans and high-resolution C 1s spectra of post-hydrothermal samples.** (A i-J i) XPS survey scans of post-hydrothermal samples. (A ii-J ii) C 1s high-resolution spectra of post-hydrothermal samples. Deconvolution of high-resolution spectra reveals the presence of sp³ hybridized carbon phase in the post-hydrothermal samples, along with carbon in sp² hybridized phase and in the functional groups. The results from this figure are described in detail in the main manuscript.

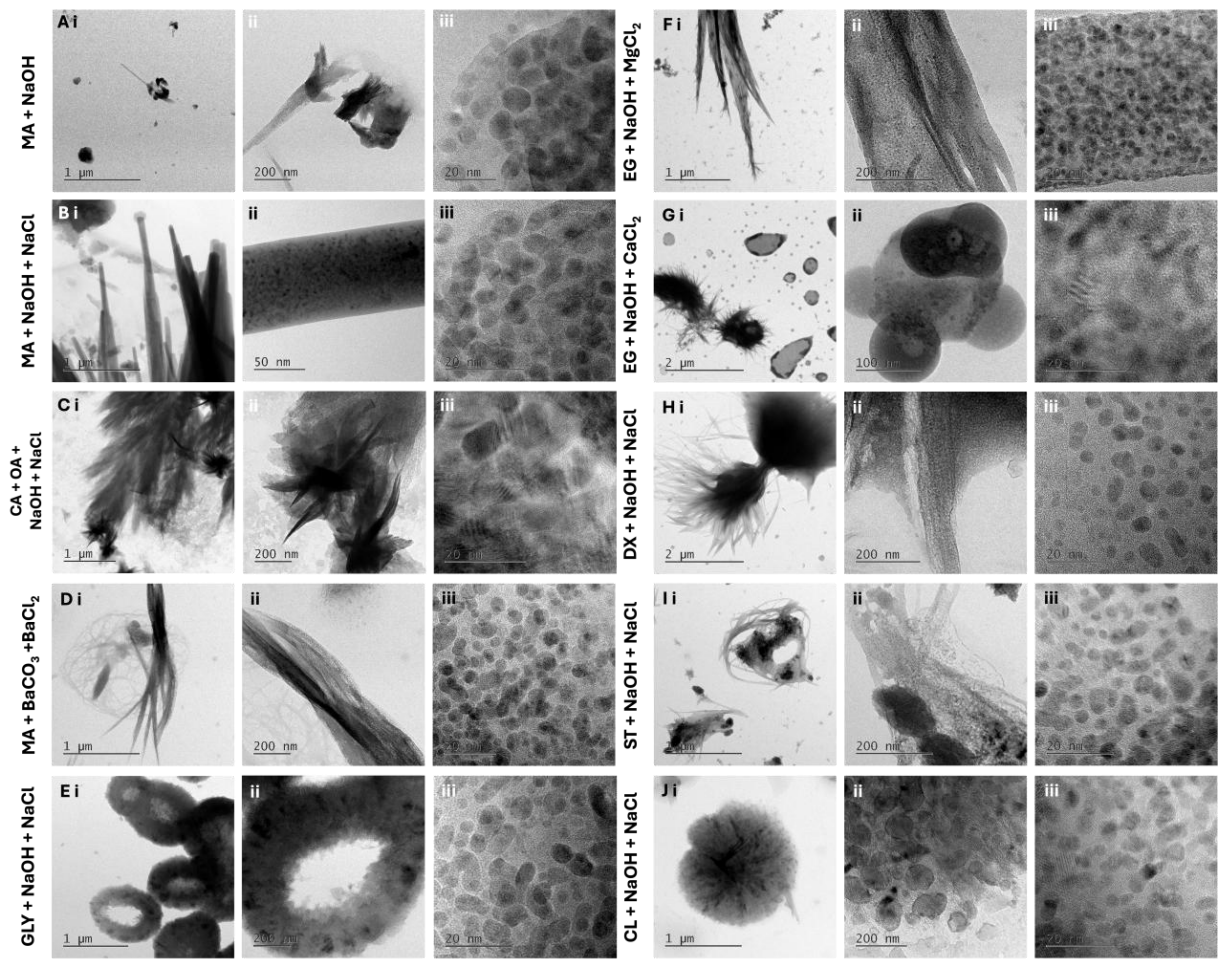

**Figure S2. Particle size distribution of nanodiamonds in post-hydrothermal samples.** (A) MA+NaOH+NaCl. (B) MA+BaCO$_3$+MA+BaCl$_2$. (C) MA+NaOH. (D) CA+OA+NaOH+NaCl. (E) GLY+NaOH+NaCl. (F) EG+NaOH+CaCl$_2$. (G) EG+NaOH+MgCl$_2$. (H) DX+NaOH+NaCl. (I) CL+NaOH+NaCl. (J) ST+NaOH+NaCl. The dried post-hydrothermal samples contain distinct morphological features at lower magnifications, including spikes and broom-like structures, spherical, annular, and random micro-agglomerates (A i-J i). Upon increasing magnification, distinct nanocrystalline phases embedded within the amorphous matrix were visible (A iii-J iii). The matrix could consist of self-assembled amorphous carbon, partially graphitized material, or residual polymerized carbonized intermediates influenced by residual ionic species, necessitating further characterization. The agglomeration could be formed by dissolving the post-hydrothermal samples in an organic solvent, followed by drying on a TEM grid, where the ionic residues from the salt systems may act as structure-directing agents, facilitating the aggregation of carbonaceous products.

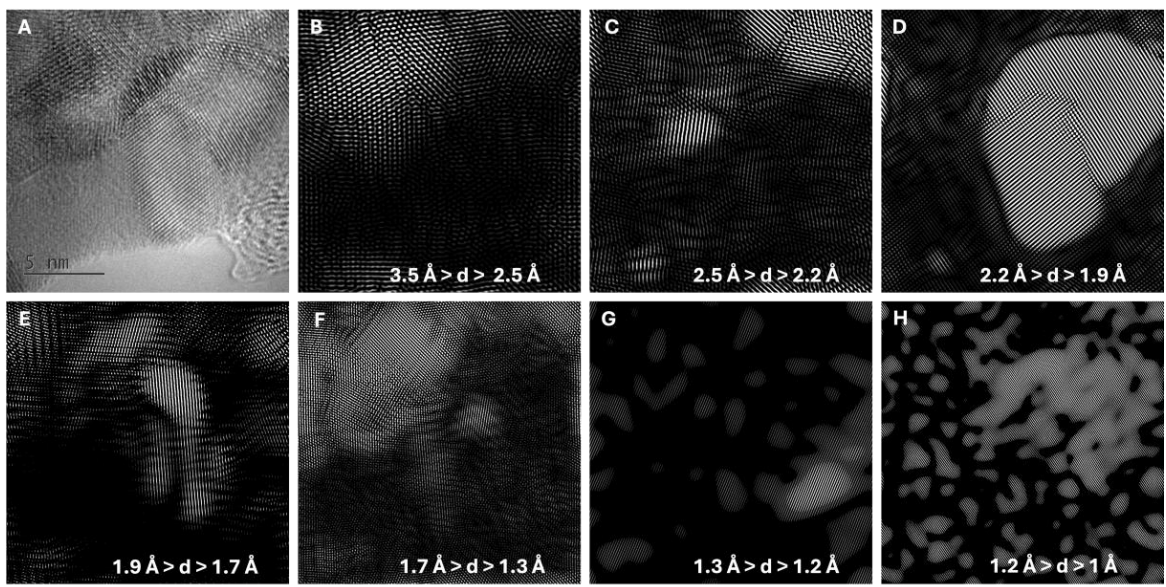

**Figure S3: Spatial distribution of crystalline lattice corresponding to different d spacings revealed from ultra-HRTEM of two fused nanodiamond crystals from CA+OA+NaOH+NaCl hydrothermal chemistry.** (A) Ultra-HRTEM of two fused nanodiamond crystals nucleating from the graphitic phase. (B) Crystalline lattices corresponding to d spacings within 3.5-2.5 Å. (C) Crystalline lattices corresponding to d spacings within 2.5-2.2 Å. (D) Crystalline lattices corresponding to d spacings within 2.2-1.9 Å. (E) Crystalline lattices corresponding to d spacings within 1.9-1.7 Å. (F) Crystalline lattices corresponding to d spacings within 1.7-1.3 Å. (G) Crystalline lattices corresponding to d spacings within 1.3-1.2 Å. (H) Crystalline lattices corresponding to d spacings within 1.2-1 Å.

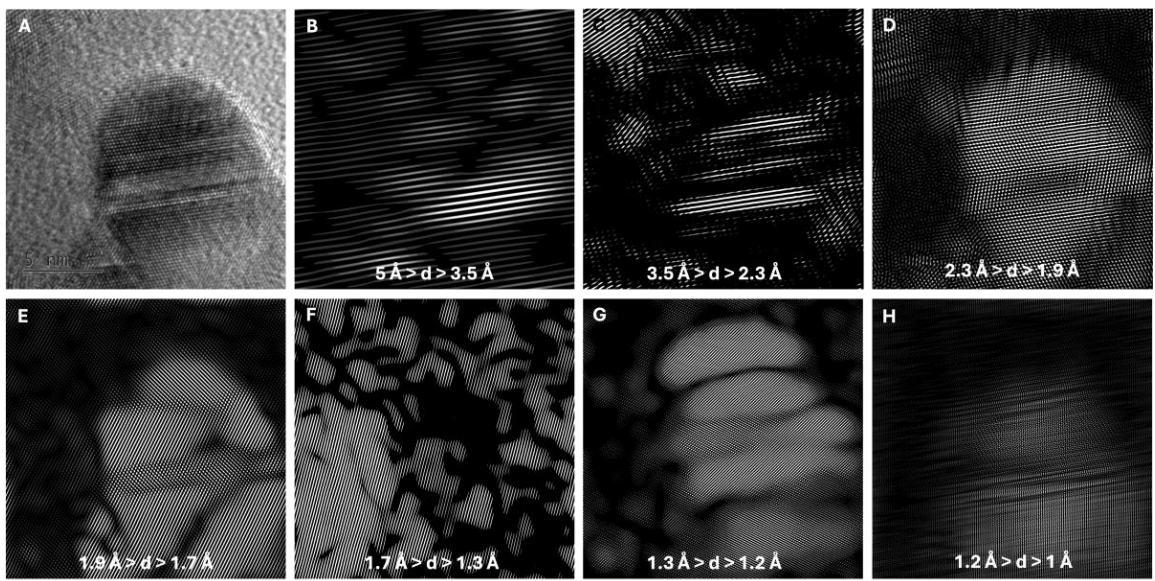

**Figure S4: Spatial distribution of crystalline lattice corresponding to different d spacings revealed from ultra-HRTEM of nanodiamond crystal from MA+BaCO₃+BaCl₂ hydrothermal chemistry.** (A) Ultra-HRTEM of the nanodiamond crystal with hierarchically stacked twins. (B) Crystalline lattices corresponding to d spacings within 5-3.5 Å. (C) Crystalline lattices corresponding to d spacings within 3.5-2.3 Å. (D) Crystalline lattices corresponding to d spacings within 2.3-1.9 Å. (E) Crystalline lattices corresponding to d spacings within 1.9-1.7 Å. (F) Crystalline lattices corresponding to d spacings within 1.7-1.3 Å. (G) Crystalline lattices corresponding to d spacings within 1.3-1.2 Å. (H) Crystalline lattices corresponding to d spacings within 1.2-1 Å.

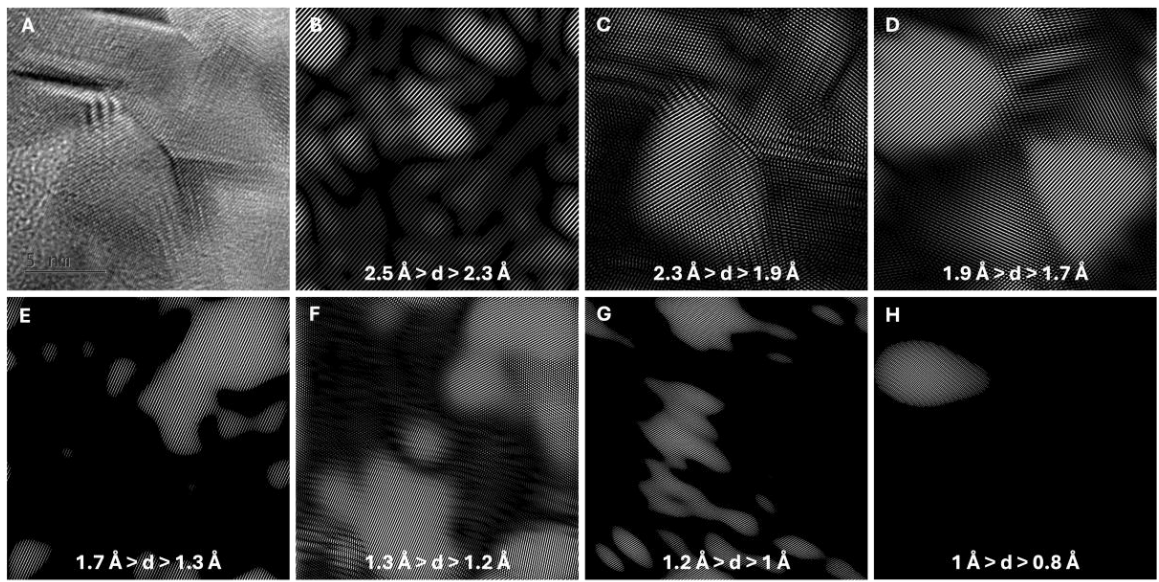

**Figure S5: Spatial distribution of crystalline lattice corresponding to different d spacings revealed from ultra-HRTEM of nanodiamond crystal from MA+BaCO₃+BaCl₂ hydrothermal chemistry.** (A) Ultra-HRTEM of the nanodiamond crystal with hierarchically stacked twins. (B) Crystalline lattices corresponding to d spacings within 2.5-2.3 Å. (C) Crystalline lattices corresponding to d spacings within 2.3-1.9 Å. (D) Crystalline lattices corresponding to d spacings within 1.9-1.7 Å. (E) Crystalline lattices corresponding to d spacings within 1.7-1.3 Å. (F) Crystalline lattices corresponding to d spacings within 1.3-1.2 Å. (G) Crystalline lattices corresponding to d spacings within 1.2-1 Å. (H) Crystalline lattices corresponding to d spacings within 1-0.8 Å.

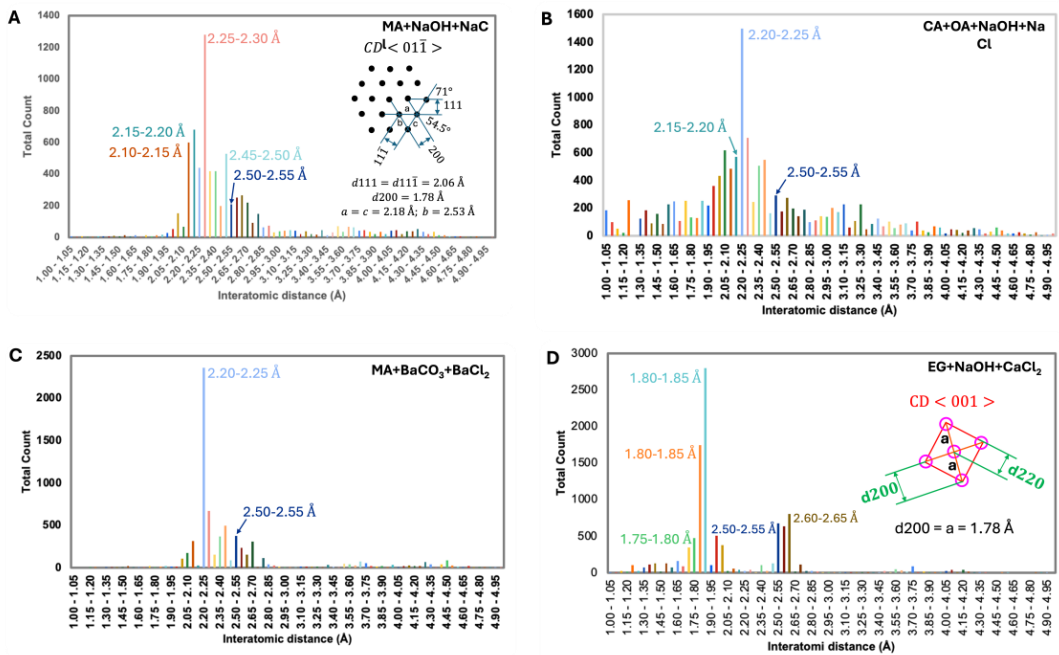

**Figure S6: Distribution of interatomic distances in the synthesized nanodiamond crystal deduced from ultra-HRTEM images.** (A) The distribution of interatomic distances in nanodiamond crystals found in post-hydrothermal samples from MA+NaOH+NaCl chemistry (Figure 2). The crystal includes $< 01\bar{1} >$ rotation and reflection twins. On $< 01\bar{1} >$ face, the atoms in the cubic diamond crystal possess interatomic distances of a = c = 2.18 Å and b = 2.53 Å with their neighboring atoms, which are present within the range of 2.15-2.20 Å (teal green) and 2.50-2.55 Å (blue) in the measured interatomic distances in (A). However, the distribution of interatomic distances is centered within the 2.25-2.3 Å range, and spans from 1.90-1.95 Å to 2.95-3.00 Å, suggesting inherent disorder in the crystal structure of the synthesized nanocrystal. (B) The distribution of interatomic distances in nanodiamond crystals found in post-hydrothermal samples from CA+OA+NaOH+NaCl chemistry (Figure 3). Similar to the distribution of interatomic distances in (A), the deviation and distribution of interatomic distances from a = c = 2.18Å and b = 2.53 Å suggest inherent disorder in the synthesized crystal. (C) The distribution of interatomic distances in nanodiamond crystals found in post-hydrothermal samples from MA+BaCO₃+MA+BaCl₂ chemistry (Figure 4). The nanodiamond crystal possesses hierarchically stacked $< 01\bar{1} >$ reflection twins. The deviation of interatomic distances from a = c = 2.18 Å and b = 2.53 Å suggests inherent disorder in the synthesized crystal. (D) The distribution of interatomic distances in nanodiamond crystals found in post-hydrothermal samples from EG+NaOH+CaCl₂ chemistry (Figure 5). On $< 001 >$ face, the atoms in the cubic diamond crystal possess an interatomic distance of 1.78 Å with their neighboring atoms, which are present within the range of 1.75-1.80 Å (green) in the measured interatomic distances in (D). The distribution of interatomic distances centered within the 1.80-1.85 Å (teal green) range in (D). Interatomic distances from 2.50-2.55 Å to 2.60-2.65 Å originate from the region of nucleation of the crystal, having stacked lattices specific to d spacing of 2.05-2.2 Å (blue), and 1.27-1.33 Å (green) (Figure 5E).

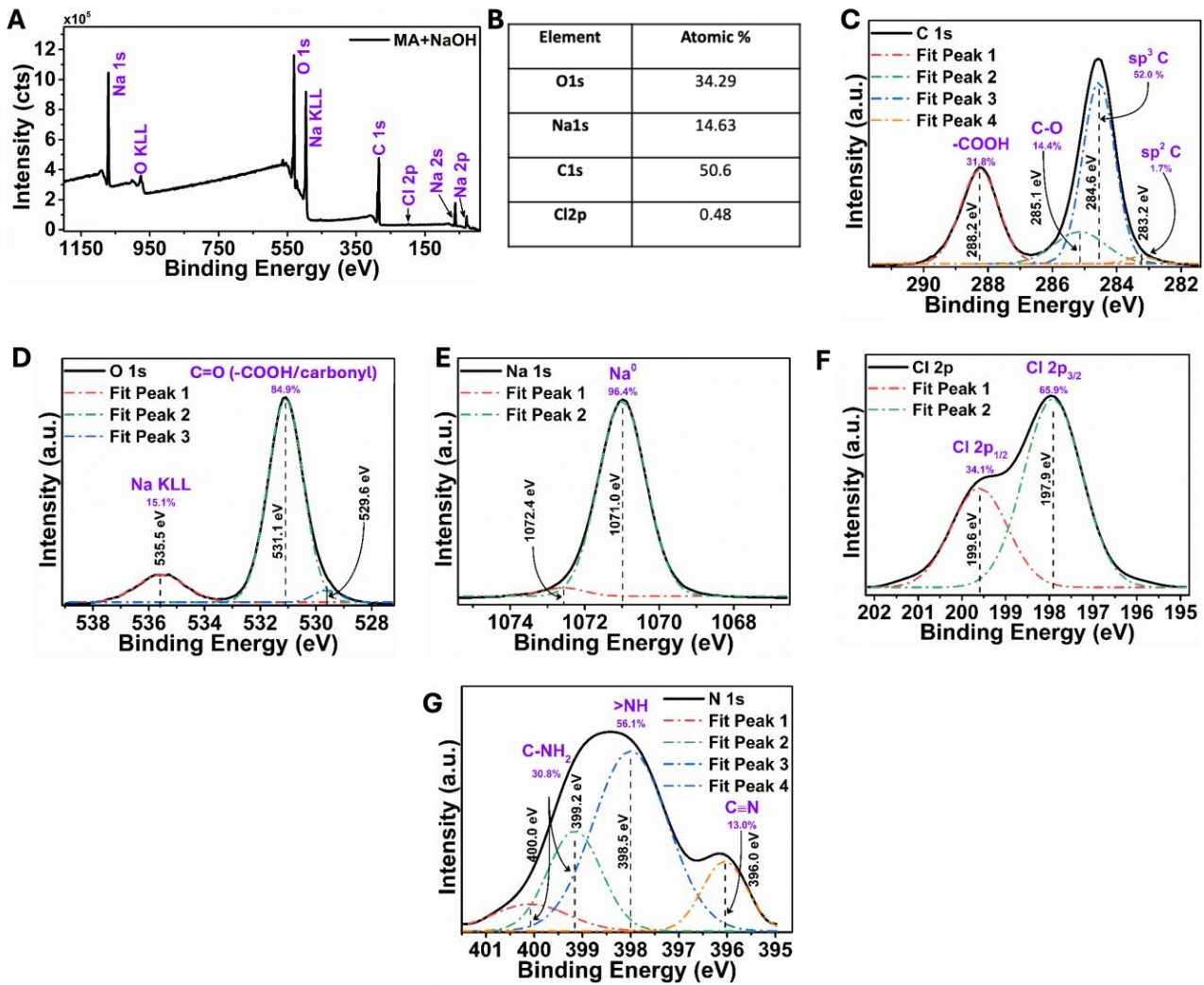

**Figure S7: XPS of post-hydrothermal samples from MA+NaOH.** (A) XPS survey scan of the post-hydrothermal sample from MA+NaOH. (B) The sample's elemental composition predominantly includes Oxygen (34.19 %), Sodium (14.63 %), Carbon (50.6 %), and Chlorine (0.48 %). (C) C 1s high- resolution spectra of post-hydrothermal samples. Deconvolution of high-resolution spectra reveals the presence of sp$^3$ hybridized carbon phase (52.0 %, 284.6 eV) in the post-hydrothermal samples, along with carbon in sp$^2$ hybridized phase (1.7 %, 283.2 eV) and in the functional groups such as -COOH, -C-O. (D) O 1s high-resolution spectra. The deconvolution of the O 1s spectra reveals the presence of absorbed Oxygen (531.1 eV) in the post-hydrothermal samples. (E) Na 1s high-resolution spectra. (F) Cl 2p high-resolution spectra. (G) Deconvolution of N 1s high-resolution spectra revealed the presence of nitrogen in the functional groups such as C-NH$_2$, -C≡N, and NH.

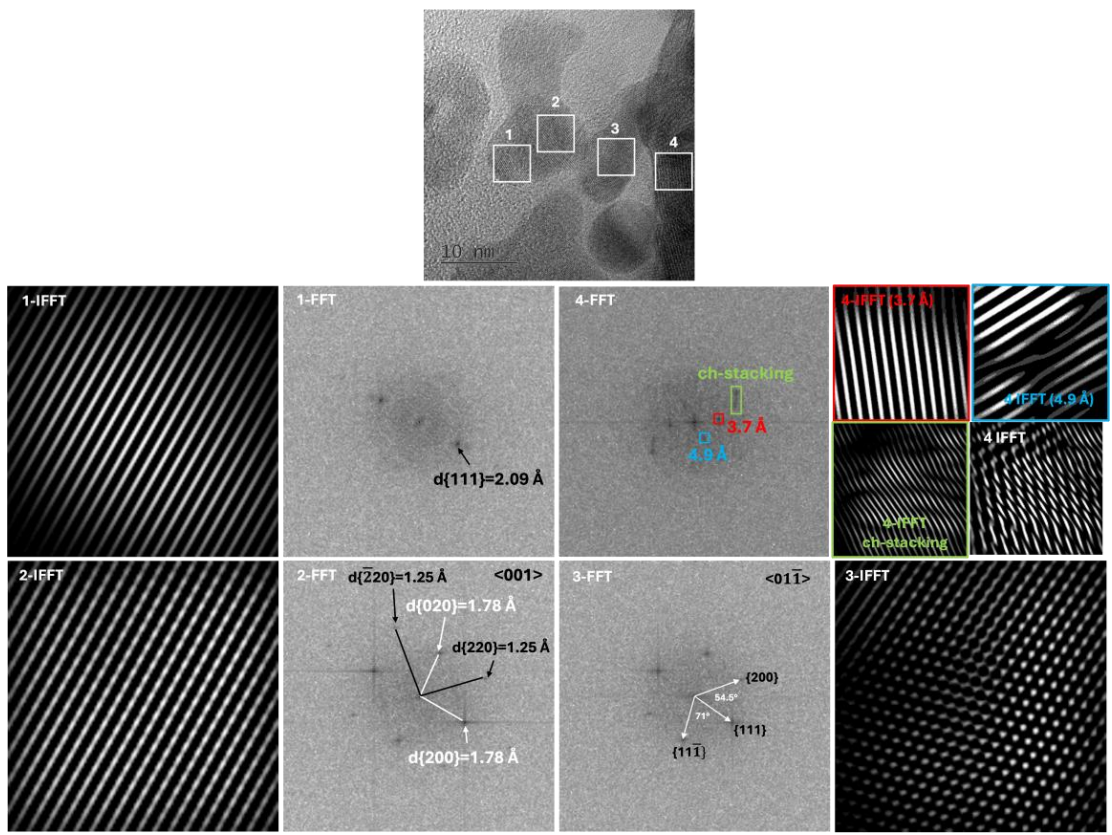

**Figure S8: HRTEM of post-hydrothermal samples from MA+NaOH chemistry.** (A) HRTEM image of the nanocrystal present in the post-hydrothermal sample. (1) FFT of region 1 reveals the presence of d1̄11 = 2.09 Å specific diamond lattice. (2) FFT of region 2 reveals the presence of d200 = 1.78 Å perpendicular to d020 = 1.78 Å and d220 = 1.25 Å perpendicular to d2̄20 = 1.25 Å crystal lattice, which is specific to ¡001¿ face of the cubic diamond crystal structure. (3) FFT of region 3 reveals the presence of ¡011̄¿ specific atomic lattice, specific to the cubic diamond phase. (4) Region 4 consists of ch-stacking disordered diamond phase.

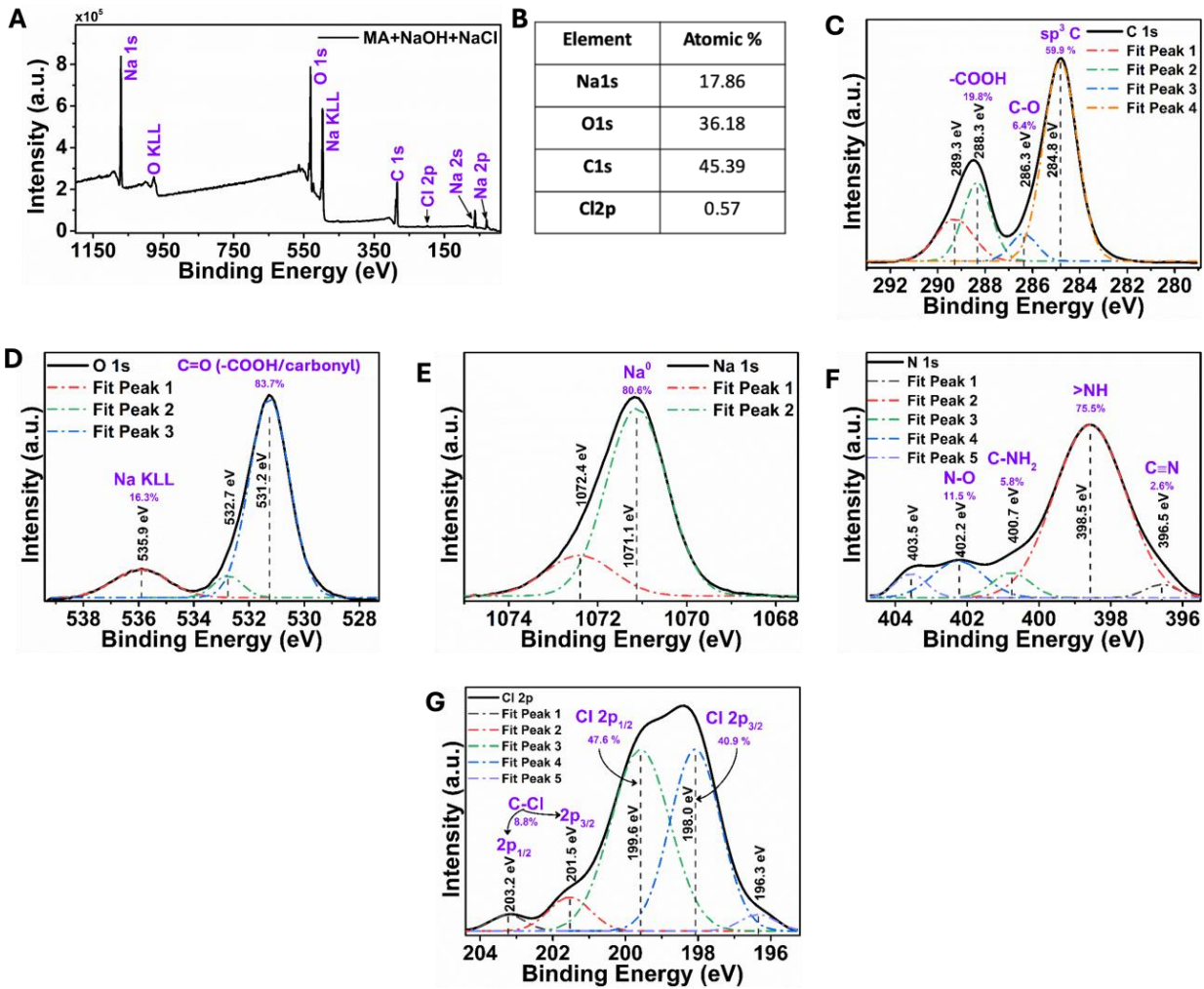

**Figure S9: XPS of post-hydrothermal samples from MA+NaOH+NaCl.** (A) XPS survey scan of the post-hydrothermal sample from MA+NaOH+NaCl. (B) The sample's elemental composition predominantly includes Sodium (17.86 %), Oxygen (36.18 %), Carbon (45.39 %), and Chlorine (0.57 %). (C) C 1s high-resolution spectra of post-hydrothermal samples. Deconvolution of high- resolution spectra reveal the presence of $sp^3$ hybridized carbon phase (284.8 eV) in the post-hydrothermal samples, along with carbon in the functional groups such as -COOH, -C-O. (D) O 1s high-resolution spectra. The deconvolution of the O 1s spectra reveals the presence of absorbed Oxygen (531.2 eV) in the post- hydrothermal samples. (E) Na 1s high-resolution spectra. (F) Deconvolution of N 1s high-resolution spectra revealed the presence of nitrogen in the functional groups such as N-O (11.5 %), C-NH$_2$ (5.8 %), -C≡N (2.6 %), and ¿NH (75.5 %). (G) Cl 2p high-resolution spectra.

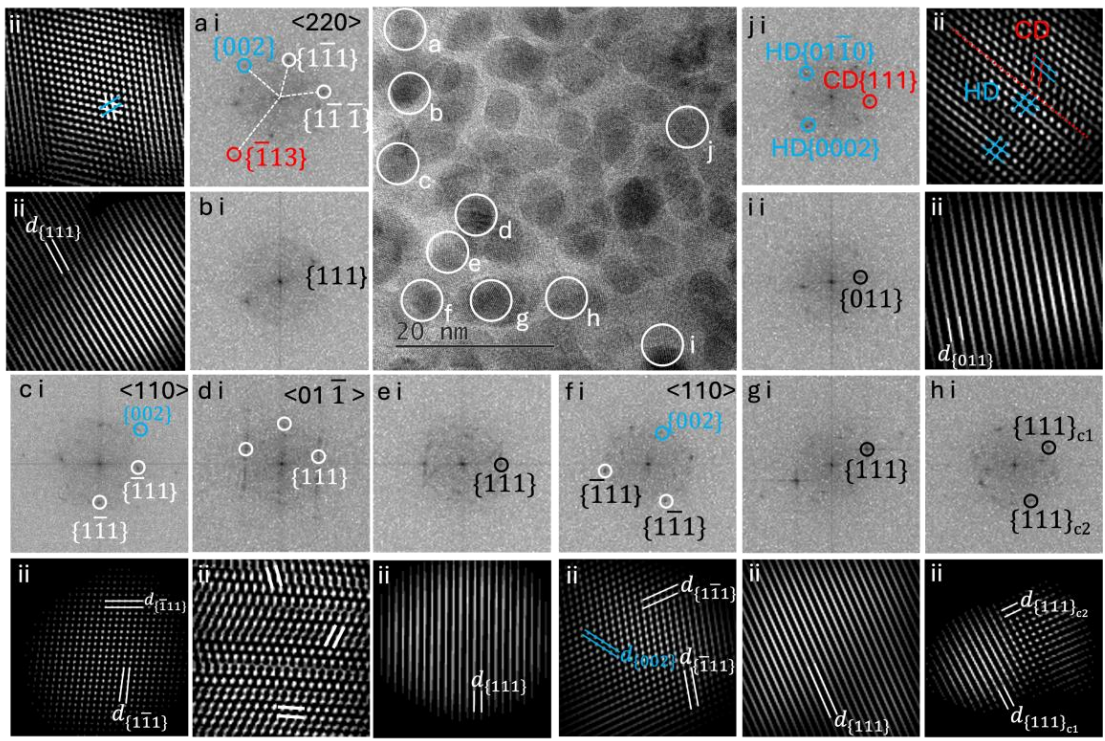

**Figure S10: HRTEM of post-hydrothermal samples from MA+NaOH+NaCl chemistry.** (A) HRTEM image of the nanocrystal present in the post-hydrothermal sample. (a) 110 projection of cubic diamond crystal structure with $1\bar{1}\bar{1}$, $1\bar{1}\bar{1}$, 002 and $\bar{1}13$ specific lattices. (b) FFT of nanocrystal reveals the presence of d111 specific diamond lattice. (c) 110 projection of cubic diamond crystal structure with $1\bar{1}\bar{1}$, $1\bar{1}\bar{1}$, and 002 specific lattices. (d) ch-stacking disordered diamond phase with $01\bar{1}$ reflections. (e) FFT of nanocrystal reveals the presence of d111 specific diamond lattice. (f) 110 projection of cubic diamond crystal structure with $1\bar{1}\bar{1}$, $1\bar{1}\bar{1}$, and 002 specific lattices. (g) FFT of nanocrystal reveals the presence of d111 specific cubic diamond lattice. (h) FFT of nanocrystal reveals perpendicularly fused d111 specific lattice of two nanodiamond crystals. (i) FFT of nanocrystal reveals the presence of d011 specific cubic diamond lattice. (j) FFT reveals the presence of an interface between the hexagonal and cubic diamond phases.

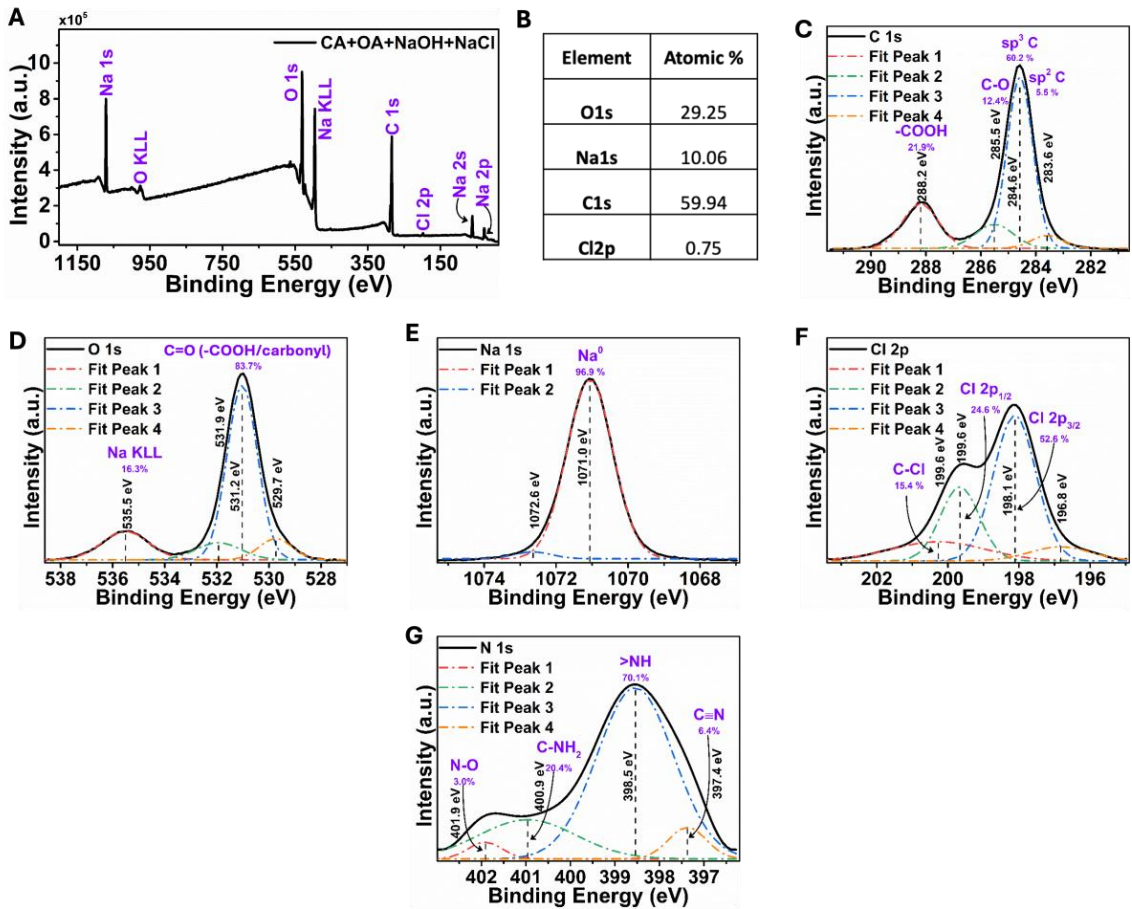

**Figure S11: XPS of post-hydrothermal samples from CA+OA+NaOH+NaCl.** (A) XPS survey scan of the post-hydrothermal sample from CA+OA+NaOH+NaCl. (B) The sample's elemental composition predominantly includes Oxygen (29.25 %), Sodium (10.06 %), Carbon (59.94 %), and Chlorine (0.75 %). (C) C 1s high-resolution spectra of post-hydrothermal samples. Deconvolution of high-resolution spectra reveals the presence of $sp^3$ hybridized carbon phase (60.2 %, 284.6 eV) in the post-hydrothermal samples, along with carbon in $sp^2$ hybridized phase (5.5 %, 283.6 eV) and in the functional groups such as -COOH, -C-O. (D) O1s high-resolution spectra. The deconvolution of the O1s spectra reveals the presence of absorbed Oxygen (531.2 eV) in the post-hydrothermal samples. (E) Na 1s high-resolution spectra. (F) Cl 2p high-resolution spectra. (G) Deconvolution of N 1s high-resolution spectra revealed the presence of nitrogen in the functional groups such as N-O (3 %), C-NH₂ (20.4 %), -C≡N (6.4 %), and -NH (70.1 %).

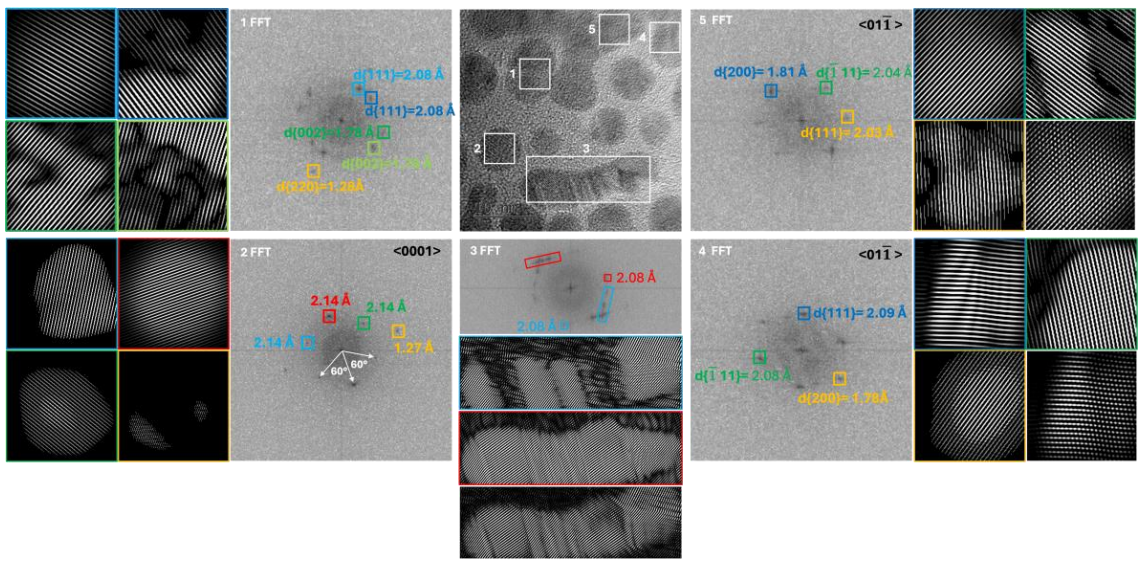

**Figure S12: HRTEM of nanodiamond synthesized CA+OA+NaOH+NaCl.** (A) HRTEM image of the nanocrystal present in the post-hydrothermal sample. (1) FFT of the region 1 reveals two fusing nanodiamond crystals with d111, d002, and d220 specific cubic diamond lattices. (2) FFT of the nanocrystal reveals atomic lattices specific to the <0001> face of graphitic lattices. (3) FFT of the region 3 shows the presence of ch-stacking disordered diamond phase. (4-5) FFT of region 4 reveals the presence of atomic lattices specific to the <01 1̄> face of the cubic diamond.

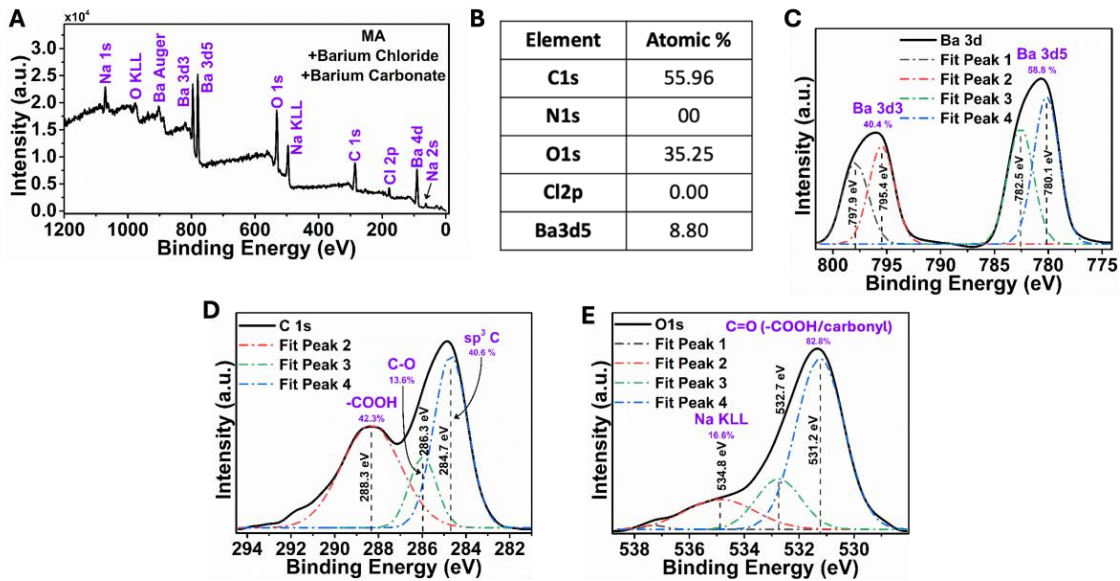

**Figure S13: XPS of post-hydrothermal samples from MA+BaCl₂+BaCO₃.** (A) XPS survey scan of the post-hydrothermal sample from MA+BaCl₂+BaCO₃. (B) The sample's elemental composition predominantly includes Carbon (55.96 %), Oxygen (35.25 %), and Barium (8.80 %). (C) Ba 3d high-resolution spectra. (D) C 1s high-resolution spectra of post-hydrothermal samples. Deconvolution of high-resolution spectra reveals the presence of $sp^3$ hybridized carbon phase (40.6 %, 284.7 eV) in the post-hydrothermal samples, along with carbon in the functional groups such as -COOH, -C-O. (E) O 1s high-resolution spectra. The deconvolution of the O 1s spectra reveals the presence of absorbed Oxygen (531.2 eV) in the post-hydrothermal samples.

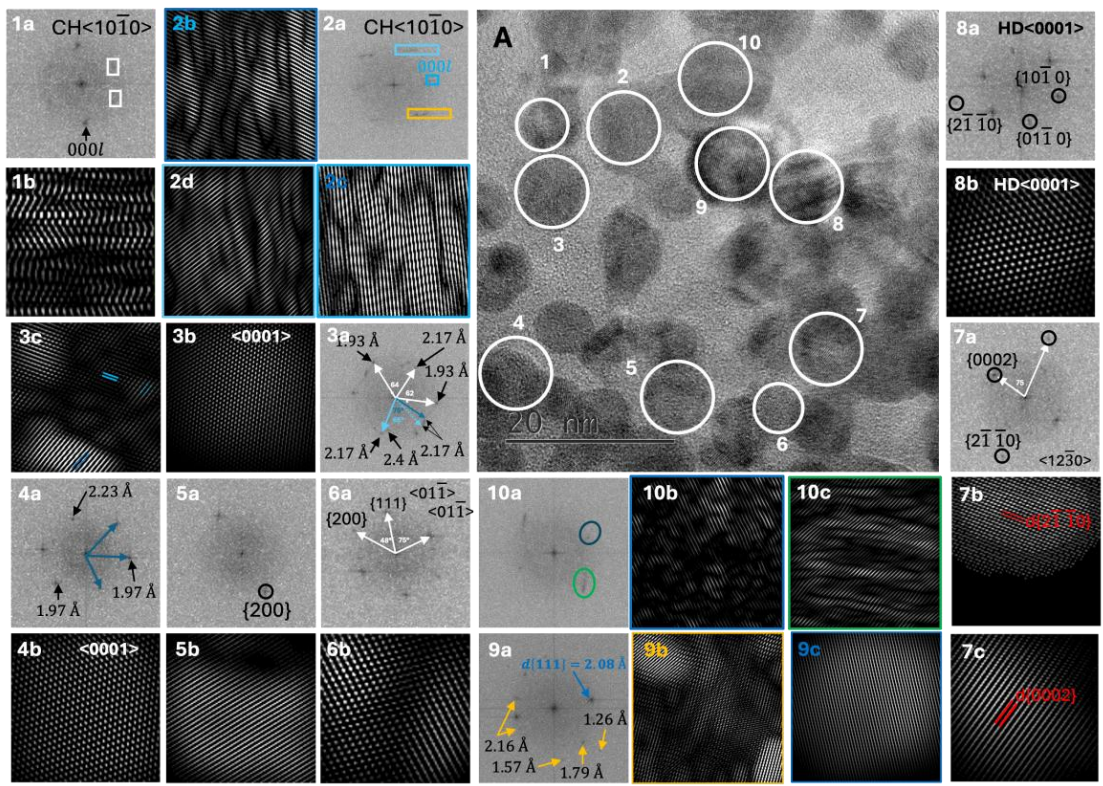

**Figure S14: HRTEM of post-hydrothermal samples from MA+BaCO₃+BaCl₂.** (A) HRTEM image of the nanocrystal present in the post-hydrothermal sample. (1a) FFT reveals the $10\bar{1}0$ projection of ch-stacking disordered diamond. (1b) Disorder can be seen in the IFFT image derived from 000l and selected spots (white rectangle) of the FFT image in (1a). (2a) FFT reveals the ¡$10\bar{1}0$¿ projection of ch-stacking disordered diamond. (2 b-d) Disorder can be seen in the IFFT image derived from 000l and selected spots of the FFT image in (2a). (3a) FFT reveals the presence of ¡0001¿ projection, which can be specific to hexagonal diamond or the graphitic phase. (3b) IFFT image shows atomic lattices specific to the ¡0001¿ projection of the graphitic lattice derived from selected spots marked with white arrows. (3c) Atomic lattices in the IFFT image derived from selected spots marked with blue arrows. (4) FFT-IFFT filtering reveals the presence of <0001> specific atomic lattices with deviated d spacing values 2.23 Å, 1.97 Å, and 1.97 Å - that can be specific to distorted hexagonal diamond or the graphitic crystal structure. (5) FFT-IFFT filtering reveals atomic lattices specific to 002. (6) FFT-IFFT reveals the presence of <$01\bar{1}$> projection specific to cubic diamond crystal. (7) FFT-IFFT reveals <$12\bar{3}0$> projection specific to hexagonal diamond lattice. (8a) FFT reveals <0001> specific projection of hexagonal diamond crystal structure. (8b) IFFT image shows the hexagonal atomic lattices specific to <0001> projection of the hexagonal diamond crystal structure. (9) FFT-IFFT filtering reveals region-9 predominantly consists of 111 specific diamond crystal lattices. (10) FFT-IFFT reveals the presence of a stacking disordered diamond phase.

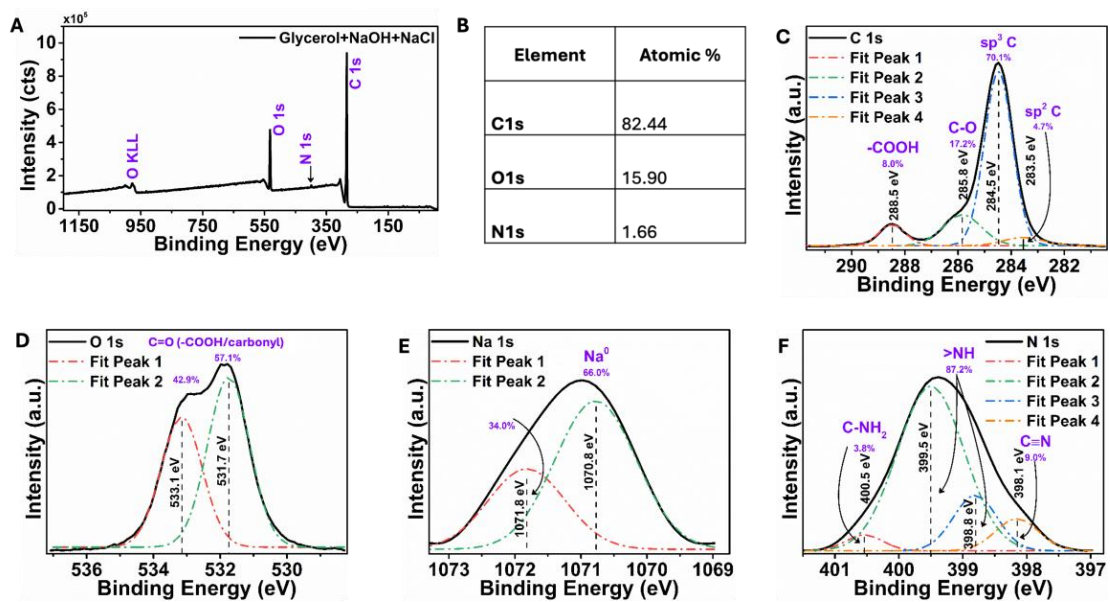

**Figure S15: XPS of post-hydrothermal samples from Glycerol (GLY)+NaOH+NaCl.** (A) XPS survey scan of the post-hydrothermal sample from GLY+NaOH+NaCl. (B) The sample's elemental composition predominantly includes Carbon (82.44 %), Oxygen (15.90 %), Carbon (59.94 %), and Nitrogen (1.66 %). (C) C 1s high-resolution spectra of post-hydrothermal samples. Deconvolution of high-resolution spectra reveals the presence of $sp^3$ hybridized carbon phase (70.1 %, 284.5 eV) in the post-hydrothermal samples, along with carbon in $sp^2$ hybridized phase (4.7 %, 283.5 eV) and in the functional groups such as -COOH, -C-O. (D) O 1s high-resolution spectra. The deconvolution of the O 1s spectra reveals the presence of absorbed Oxygen (531.7 eV) in the post-hydrothermal samples. (E) Na 1s high-resolution spectra. (F) Deconvolution of N 1s high-resolution spectra revealed the presence of nitrogen in the functional groups such as C-NH$_2$ (3.8 %), -C≡N (9 %), and NH (87.2 %).

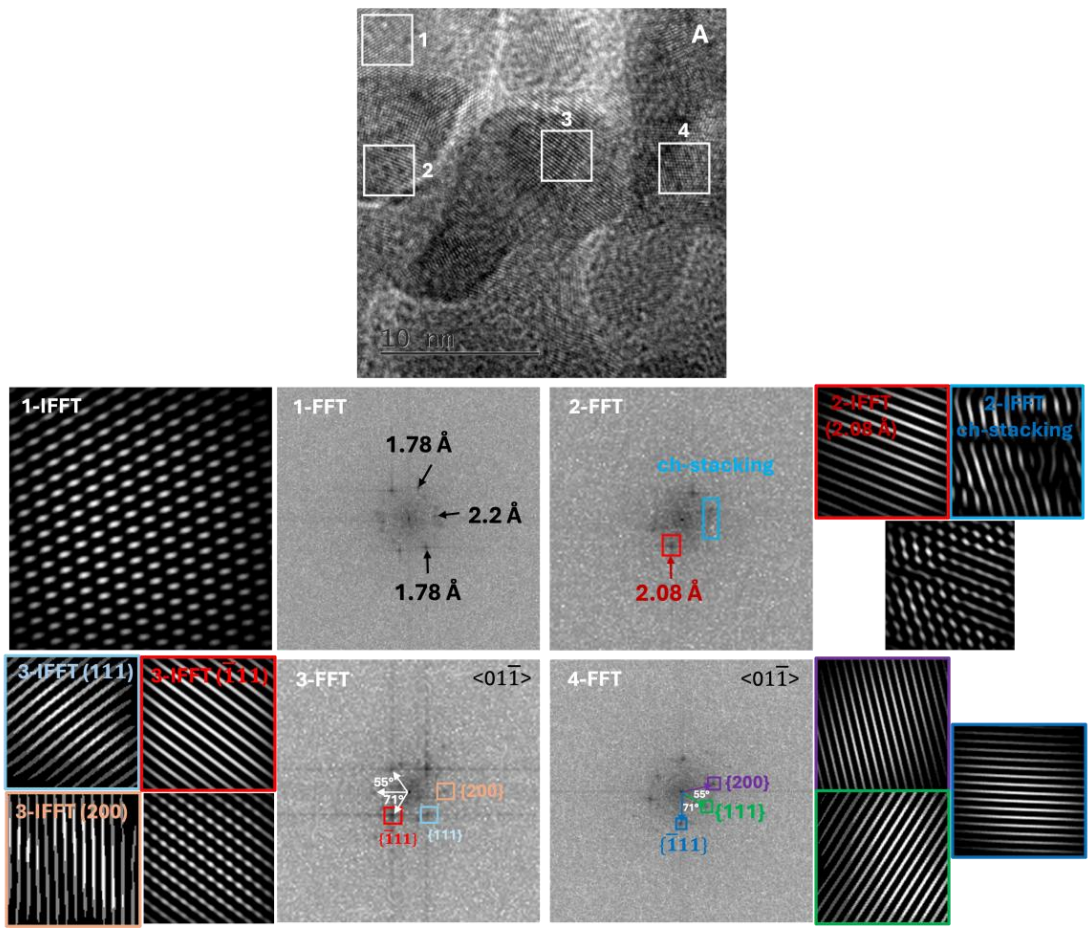

**Figure S16: HRTEM of the post-hydrothermal samples from GLY+NaOH+NaCl.** (A) HRTEM image of the nanocrystal present in the post-hydrothermal sample. (1) FFT-IFFT filtering reveals the presence of a hexagonal atomic lattice similar to distorted <01$\bar{1}$> of the cubic diamond crystal structure, where the d spacings and angle among the crystal lattices. (2) FFT-IFFT filtering reveals the presence of ch-stacking disordered diamond phase in region 2. (3-4) FFT-IFFT filtering reveals the presence of a hexagonal atomic lattice similar to <01$\bar{1}$> of the cubic diamond crystal structure.

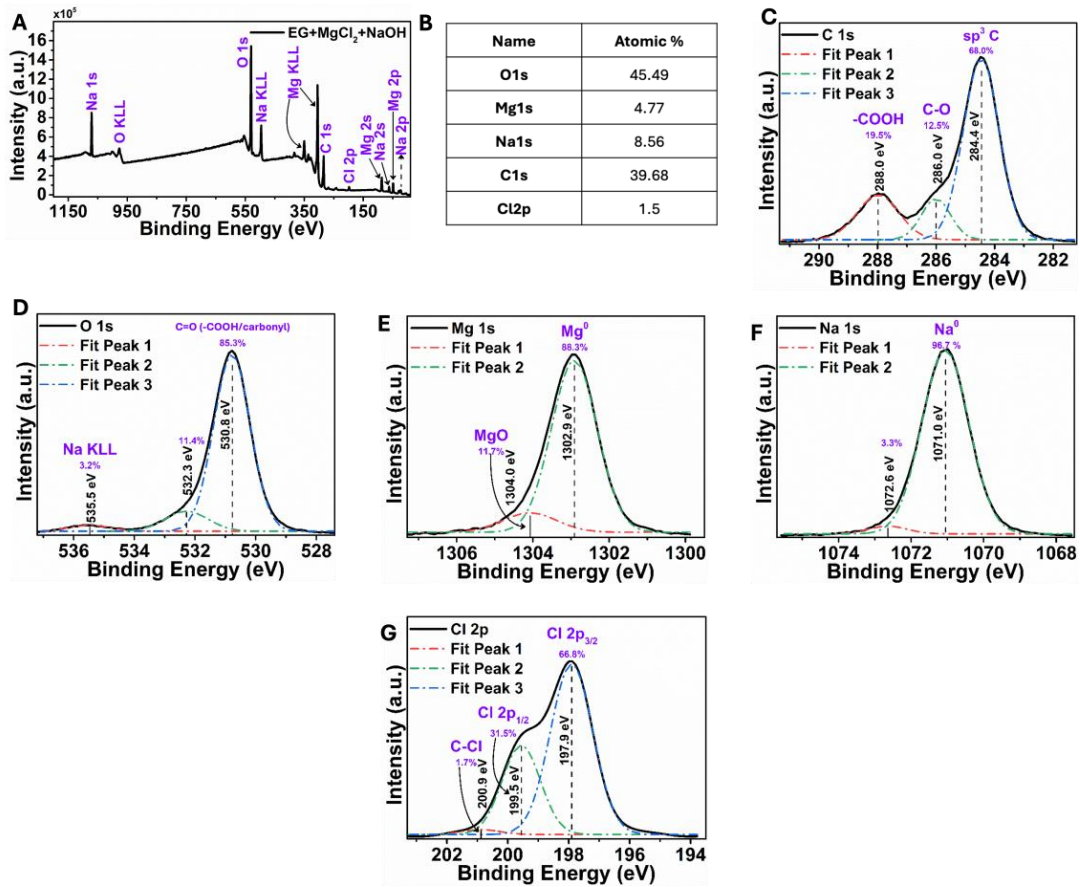

**Figure S17: XPS of post-hydrothermal samples from EG+NaOH+MgCl₂.** (A) XPS survey scan of the post-hydrothermal sample from EG+NaOH+MgCl₂. (B) The sample's elemental composition predominantly includes Oxygen (45.49 %), Magnesium (4.77 %), Sodium (8.56 %), Carbon (39.68 %), and Chlorine (1.5 %). (C) C 1s high-resolution spectra of post-hydrothermal samples. Deconvolution of high-resolution spectra reveals the presence of sp³ hybridized carbon phase (68.0 %, 284.4 eV) in the post-hydrothermal samples, along with carbon in the functional groups such as -COOH, -C-O. (D) O 1s high-resolution spectra. The deconvolution of the O 1s spectra reveals the presence of absorbed Oxygen (530.8 eV) in the post-hydrothermal samples. (E) Mg 1s high-resolution spectra. (F) Na 1s high-resolution spectra. (G) Cl 2p high-resolution spectra.

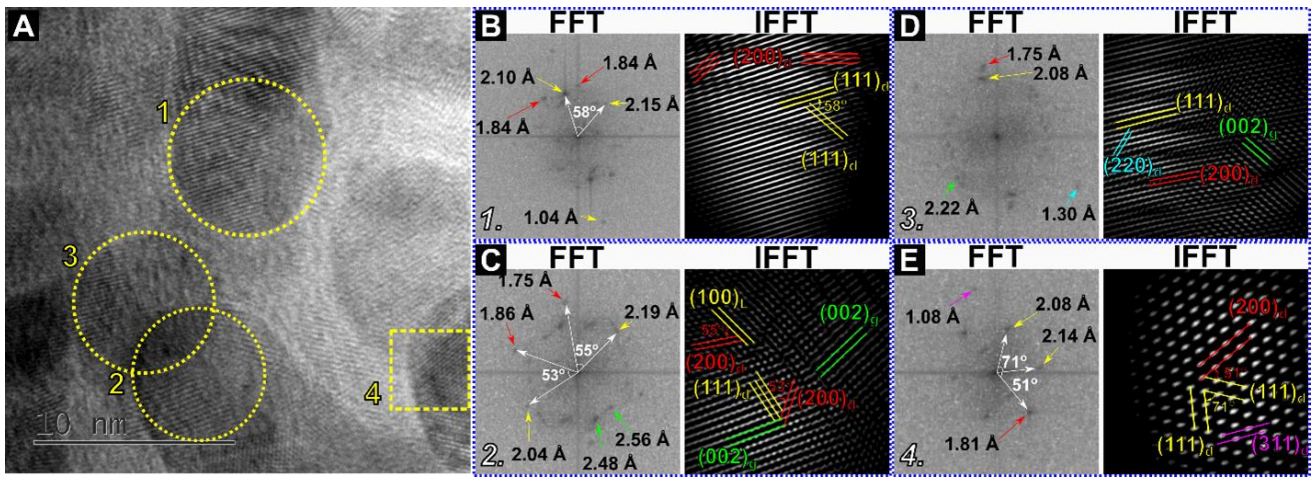

**Figure S18: HRTEM of post-hydrothermal samples from EG+NaOH+MgCl₂.** FFT analysis of particle (A) was undertaken to determine the atomic arrangement in the crystalline phases. (B) Particle 1 represents a crystalline nanodiamond, confirmed by the atomic arrangement in lattice d-spacing of 2.15- 2.10 Å and 1.84 Å, while the interplanar angle of 58° composed by (1 1 1) planes indicate the stacking fault in the crystal. (C)Particle 2 is composed of two overlapping nanodiamond crystals and the formation of a graphitic carbon phase at the surface. The lattice d-spacings of 2.19 Å and 2.04 Å represent the (1 1 1) plane, while 1.86-1.75 Å, represent the (2 0 0) planes of the nanodiamond phase. The interplanar angle of 55°-53° in two overlapping crystals is significant to cubic nanodiamonds. (D) Particle 3 constitute of atoms arranged in lattice d-spacings specific to nanodiamond/lonsdaleite crystals (2.22 Å, 2.08 Å, 1.75 Å, and 1.30 Å); however, the structure seems to be higher disordered, possibly due to overlapping crystal or stacking faults. (E) The lattice d-spacing of 2.14-2.08 Å and 1.81 Å in particle 4, corresponding to (1 1 1) and (2 0 0) planes and the angle composed by these planes (1 1 1) and (2 0 0) are specific to nanodiamond crystals. The smaller d-spacing of 1.08 Å corresponds to the (3 1 1) is also specific to nanodiamonds. The formation of nanodiamonds was confirmed with the FFT analysis of the nanoparticles, which are composed of lattice d-spacings of $2.19 - 2.14$ Å (orange), $2.10 - 2.04$ Å (yellow), $1.86 - 1.75$ Å (red), $1.30$ Å (pink) and $1.08 - 1.04$ Å (yellow, secondary reflection of (1 1 1)), characteristic of crystalline planes of nanodiamond polymorphs, i.e., cubic diamond and lonsdaleite, as discussed earlier. Particle 4 exhibits the perfect cubic nanodiamond, where the (1 1 1) plane composes $\bar{7}0°$angle, while the (1 1 1) and (2 0 0) planes compose an angle of $\bar{5}5°$, matching the theoretical interplanar angle in the diamond crystal. Particles 2 and 3 report the formation of graphitic carbon, confirmed by the lattice d-spacing above 2.22Å (green).

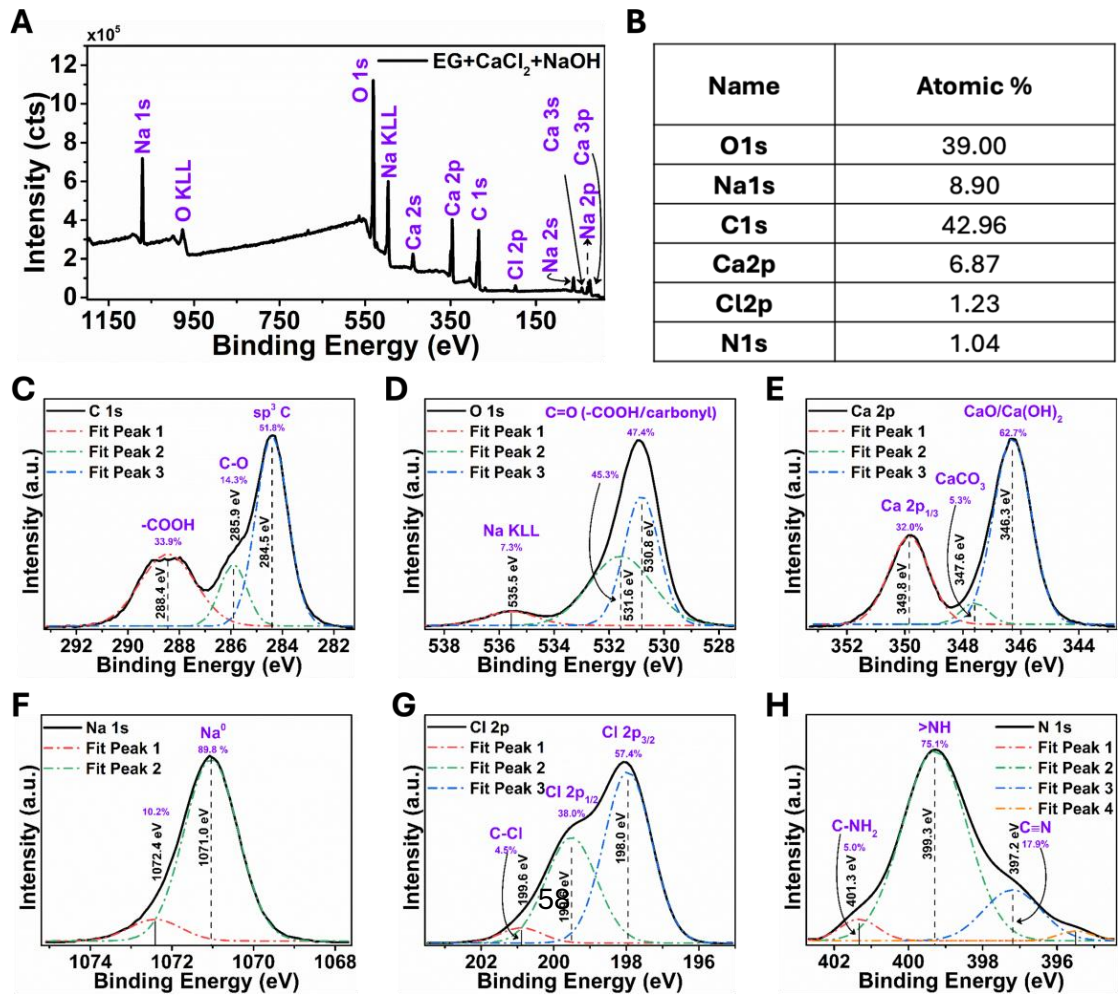

**Figure S19: XPS of post-hydrothermal samples from EG+NaOH+CaCl₂.** (A) XPS survey scan of the post-hydrothermal sample from EG+NaOH+CaCl₂. (B) The sample's elemental composition predominantly includes Oxygen (39.00 %), Sodium (8.90 %), Carbon (42.96 %), Calcium (6.87 %),

Chlorine (1.23 %), and Nitrogen (1.04 %). (C) C 1s high-resolution spectra of post-hydrothermal samples. Deconvolution of high-resolution spectra reveals the presence of sp³ hybridized carbon phase (51.8 %, 284.5 eV) in the post-hydrothermal samples, along with carbon in the functional groups such as -COOH,

-C-O. (D) O 1s high-resolution spectra. The deconvolution of the O 1s spectra reveals the presence of absorbed Oxygen (530.8 eV) in the post-hydrothermal samples. (E) Deconvolution of Ca 2p high- resolution spectra reveals the presence of Calcium in the form of hydrated Calcium oxide or Calcium hydroxide (CaO/Ca(OH)₂, 346.3 eV, 62.7 %) and Calcium carbonate (CaCO₃, 347.6 eV, 5.3 %). (F) Na 1s high-resolution spectra. (G) Cl 2p high-resolution spectra. (H) Deconvolution of N 1s high-resolution spectra revealed the presence of nitrogen in the functional groups such as C-NH₂ (5.0 %), -C≡N (17.9 %), and ¿NH (75.1 %).

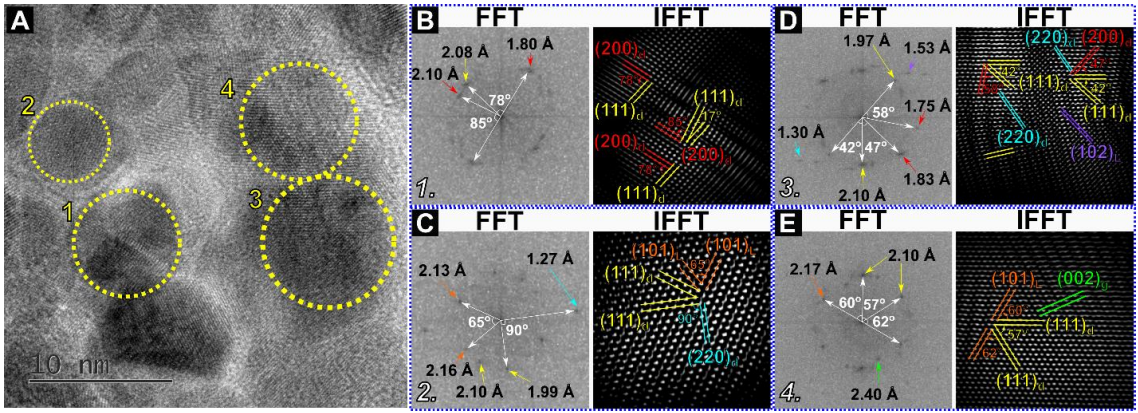

**Figure S20: HRTEM of post-hydrothermal samples from EG+NaOH+CaCl₂.** FFT analysis of the nanoparticles (A) was performed to determine the atomic arrangement and crystalline phases. (B) Particle 1 represents the crystalline nanodiamond phase, confirmed by lattice d-spacing of 2.10-2.08 Å and 1.80 Å specific to nanodiamonds (1 1 1) and (2 0 0) planes. However, the particle showcases the stacking fault with an extended interplanar angle (55° à 78°) and structural defect in the crystalline structure. (C) Particle 2 exhibits the atomic arrangement such that lattice d-spacings of 2.16-2.13 Å, 2.10-1.99 Å, and 1.27 Å are closely correlated to lonsdaleite or cubic nanodiamonds. However, the lattice arrangement represents a disordered crystal structure with stacking faults and twinning. (D) Particle 3 constitutes lattice d-spacings specific to lonsdaleite or cubic nanodiamonds. Meanwhile, the deviation from a natural interplanar angle and twinning in the structure indicate structural defects in nanodiamonds. (E) The atomic arrangement in particle 4, with a d-spacing of 2.17-2.10 Å, specifies the nanodiamond core, and 2.40 Å indicates graphitic carbon phases surrounding the nanodiamond. Particle 4 also represents the structural defect in nanodiamonds.

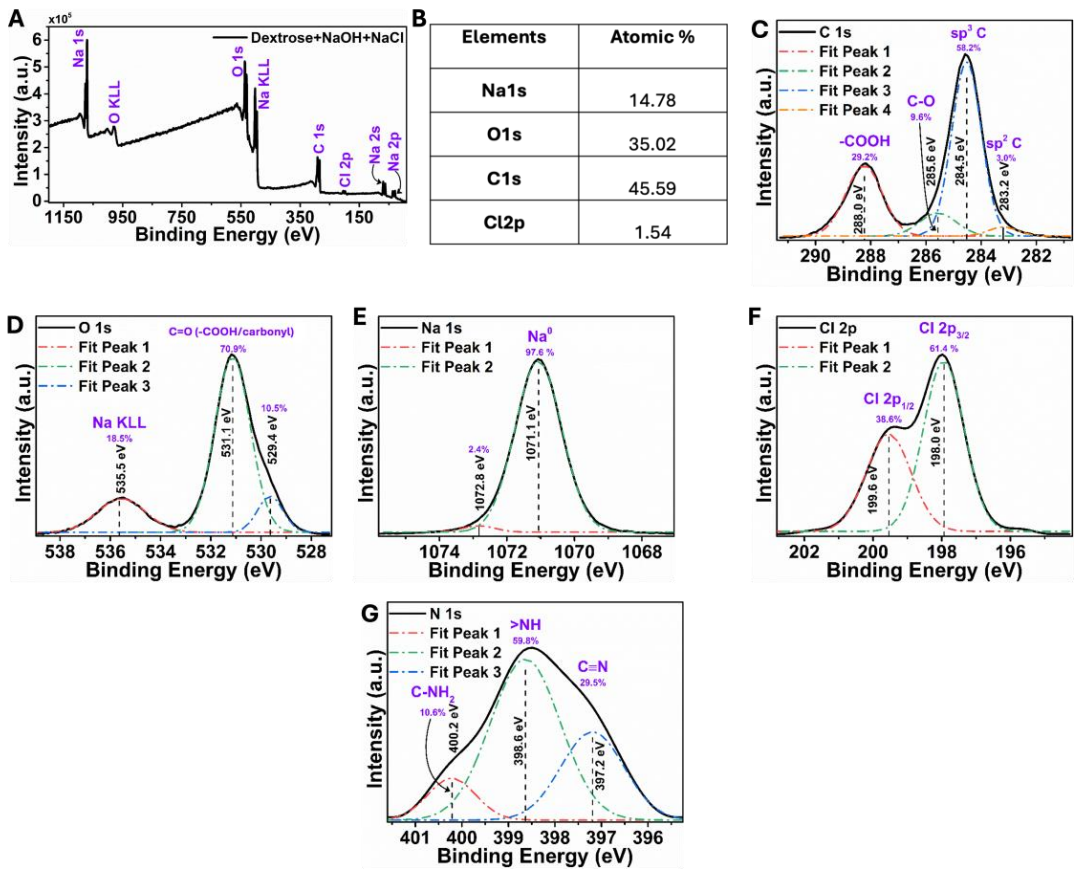

**Figure S21: XPS of post-hydrothermal samples from Dextrose (DX)+NaOH+NaCl.** (A) XPS survey scan of the post-hydrothermal sample from DX+NaOH+NaCl. (B) The sample's elemental composition predominantly includes Sodium (14.78 %), Oxygen (35.02 %), Carbon (45.59 %), and Chlorine (1.54 %). (C) C 1s high-resolution spectra of post-hydrothermal samples. Deconvolution of  high-resolution spectra reveals the presence of sp$^3$ hybridized carbon phase (58.2 %, 284.5 eV) in the post-hydrothermal samples, along with carbon in sp$^2$ hybridized carbon phase (3.0 %, 283.2 eV) and in the functional groups such as -COOH, -C-O. (D) O 1s high-resolution spectra. The deconvolution of the O 1s spectra reveals the presence of absorbed Oxygen (531.1 eV) in the post-hydrothermal samples. (E) Na 1s high-resolution spectra. (F) Cl 2p high-resolution spectra. (G) Deconvolution of N 1s high- resolution spectra revealed the presence of nitrogen in the functional groups such as C-NH$_2$ (10.6 %), -C≡N (29.5 %), and ¿NH (59.8 %).

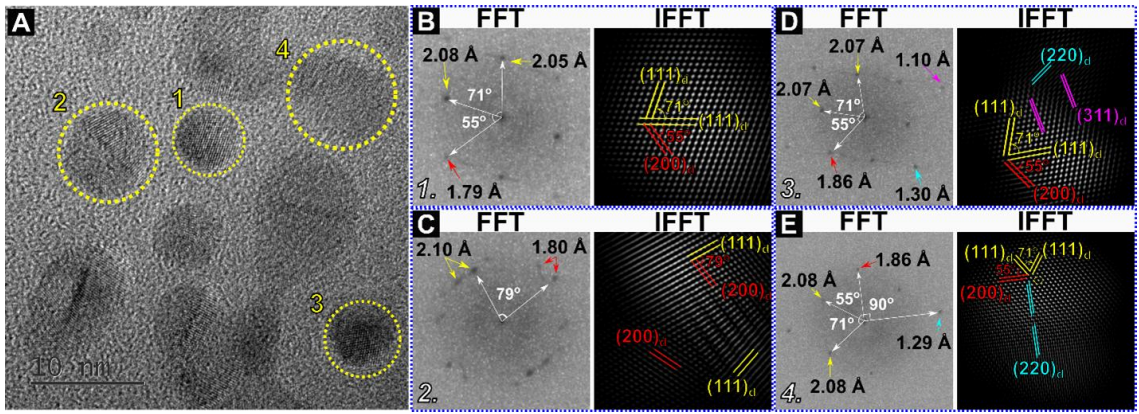

**Figure S22: HRTEM of post-hydrothermal samples from Dextrose (DX)+NaOH+NaCl.** FFT analysis of the particles (A) confirmed the nanodiamond formation. (B) Particle 1 is composed of an atomic arrangement at a lattice d-spacing of 2.05-2.08 Å and 1.79 Å, corresponding to (1 1 1) and (2 0 0) crystalline planes in nanodiamond. The interplanar angle of 71° between (1 1 1) and (1 1 1) and of 55° between (1 1 1) and (2 0 0) confirm the particle to be a nanodiamond. (C) Particle 2 is composed of lattice d-spacing corresponding to (1 1 1) and (2 0 0) planes of nanodiamond, with a stacking fault in the atomic arrangement, where the interplanar angle between two planes increases from 55° to 79°. (D) Particle 3 represents an atomic arrangement corresponding to nanodiamonds, where lattice d-spacings (2.07 Å and 1.86 Å) and interplanar angles of 71° and 55°are composed by (1 1 1) and (2 0 0) planes.

In addition, FFT analysis of particle 3 also revealed the smaller lattice d-spacing, 1.30 Å and 1.10 Å, corresponding to the (2 2 0) and (3 1 1) planes of nanodiamond. (E) Particle 4, similar to particle 3, exhibits carbon atoms arranged in a nanodiamond crystal. In particles 1, 3, and 4, the lattice d-spacings are in 2.03 – 2.10 Å and 1.80 Å, and the angle between these planes ($\overline{7}0°$ and $\overline{5}5°$), corresponding to the (1 1 1) and (2 0 0) planes of the nanodiamonds crystal lattice. Some nanodiamonds have a distorted lattice arrangement, as in particle 2, where the (1 1 1) and (2 0 0) occur at an angle of $\overline{8}0°$. Besides some lattice d-spacings, like 1.30 Å (particle 3, 4) and 1.10 Å (particle 3), are observed in nanodiamonds corresponding to (2 2 0) and (3 1 1) planes. The lattice d-spacings and the angle between the planes indicated the formation of nanodiamonds in the hydrothermal treatment of oxidized Dextrose in the alkaline medium in the presence of sodium and chloride ions.

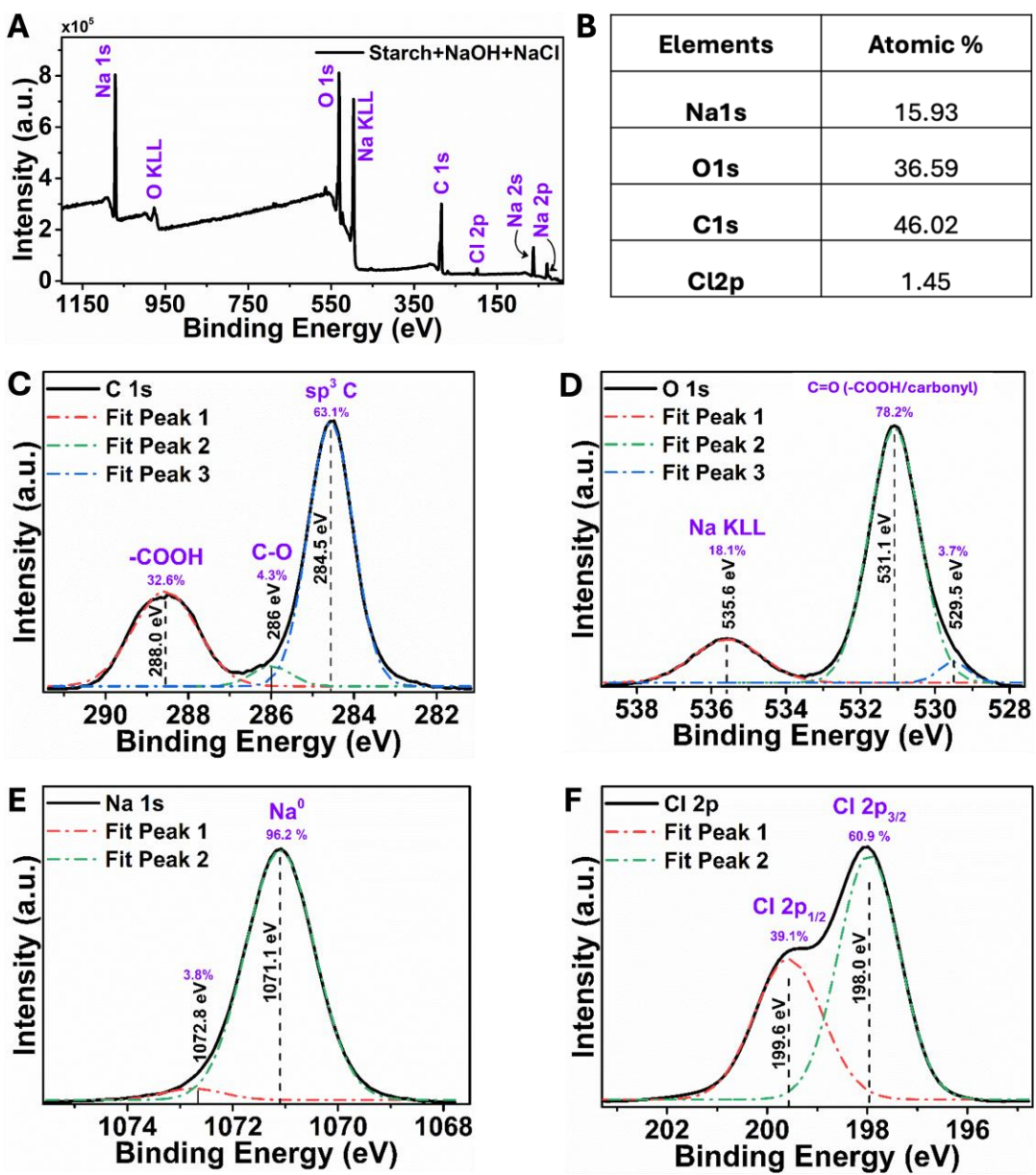

**Figure S23: XPS of post-hydrothermal samples from Starch (ST)+NaOH+NaCl.** (A) XPS survey scan of the post-hydrothermal sample from ST+NaOH+NaCl. (B) The sample's elemental composition predominantly includes Sodium (15.93 %), Oxygen (36.59 %), Carbon (46.02 %), and Chlorine (1.45%). (C) C 1s high-resolution spectra of post-hydrothermal samples. Deconvolution of high-resolution spectra reveals the presence of sp³ hybridized carbon phase (63.1 %, 284.5 eV) in the post-hydrothermal samples, along with carbon in the functional groups such as -COOH, -C-O. (D) O 1s high-resolution spectra. The deconvolution of the O 1s spectra reveals the presence of absorbed Oxygen (531.1 eV) in the post-hydrothermal samples. (E) Na 1s high-resolution spectra. (F) Cl 2p high-resolution spectra.

64

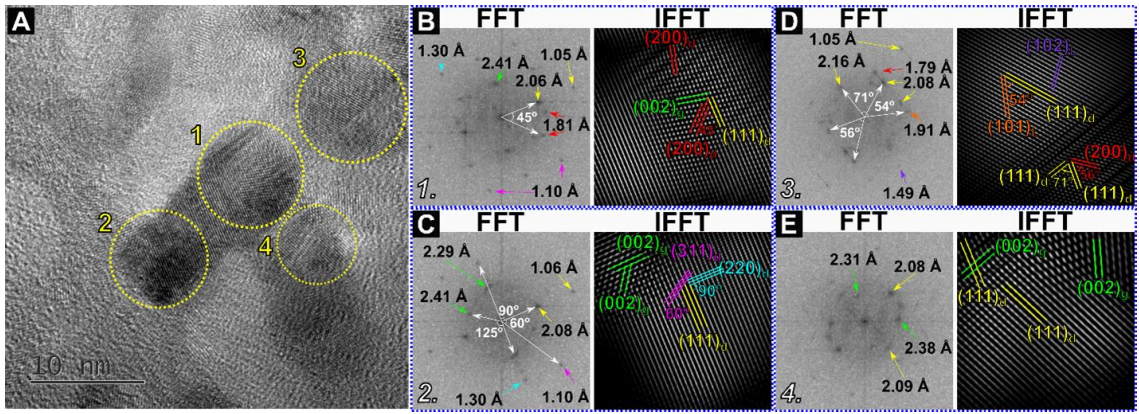

**Figure S24: HRTEM of post-hydrothermal samples from Starch (ST)+NaOH+NaCl.** (A) was performed to determine the atomic arrangement in the crystalline phases. (B) The atomic arrange- ment in particle 1, forming interlayer d-spacings of 2.05 Å, 1.81 Å, 1.30 Å, 1.10 Å, and 1.05 Å (secondary reflection of 2.05 Å), that corresponds to the nanodiamond's (1 1 1), (2 0 0), (2 2 0), (3 1 1), and (2 2 2) planes. The interlayer angle between (1 1 1) and (2 0 0) of 45° (55°) indicates the stacking fault and lattice d-spacing of 2.41 Å, indicating the formation of graphitic carbon at the nanodiamond surface. (C) The lattice d-spacings observed in particle 2 indicate nanodiamond phases. The d-spacings of 2.08 Å, 1.30 Å, 1.10 Å, and 1.05 Å (secondary reflection of 2.08 Å) correspond to nanodiamond's (1 1 1), (2 2 0), (3 1 1) and (2 2 2) planes respectively, while d-spacing of 2.41-2.29 Å, resulting from graphitic carbon phases. The absence of the crystalline plane with a d-spacing of 1.80 Å and distortion of the interplanar angle indicates the stacking fault in the nanodiamond. (D) Particle 3 exhibits atomic arrangement into lattice d-spacings of 2.16 Å, 2.08 Å, 1.91 Å, 1.79 Å, and secondary reflection of 1.05 Å, corresponding to d-spacings observed in cubic and hexagonal nanodiamonds. Both phases occurring in the same particle indicates the formation of an intricate twinned nanodiamond structure. The interplanar angles of 71° and 56° formed by these planes further indicate the formation of cubic nanodiamonds. (E) Particle 4 is dominated by lattice d-spacings of 2.38-2.31 Å and 2.09-2.08 Å, indicating the formation of both diamond and graphitic phases. The lattice d-spacings of 2.06 − 2.08 Å, 1.79 − 1.81 Å, 1.30 Å, 1.10 Å, and 1.05 −1.06 Å, distinguishes the nanoparticles as nanodiamonds, corresponding to its (1 1 1), (2 0 0), (2 2 0), (3 1 1) and (2 2 2) planes, respectively. The lattice d-spacings of 2.29 − 2.41 Å indicate the formation of different graphitic carbon phases over the nanodiamond surface. Twinning in the nanodiamond crystals was observed in particle 3, where the nanodiamonds are composed of lattice d-spacings of 2.16 Å and 1.91 Å, which are consistent with the lonsdaleite diamond polymorph, corresponding to (1 0 0) and (1 0 1) planes. However, the orientation of planes was distorted (except particle 3), which was possible due to the overlapping of nanodiamond crystals.

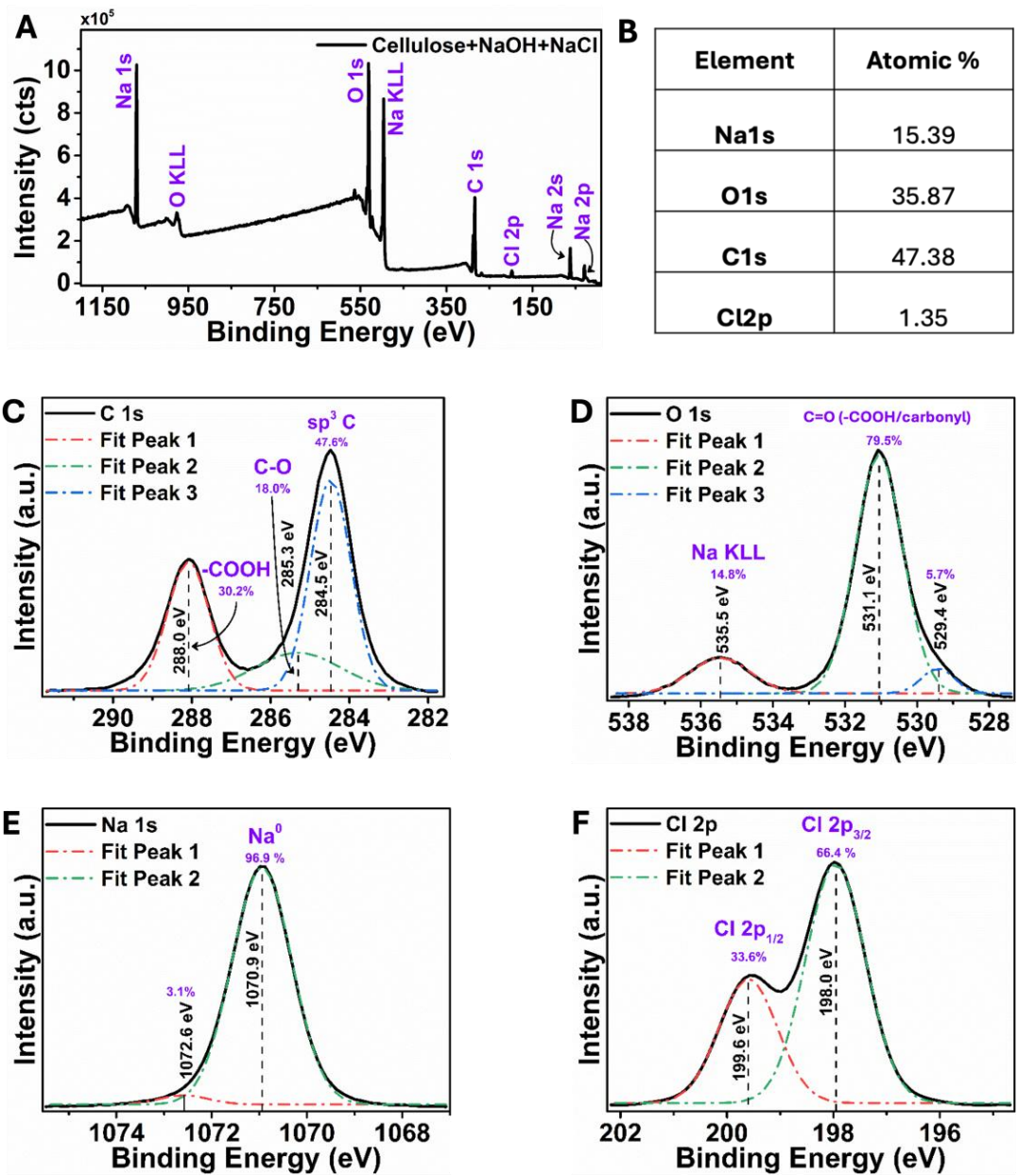

**Figure S25: XPS of post-hydrothermal samples from Cellulose (CL)+NaOH+NaCl.** (A) XPS survey scan of the post-hydrothermal sample from CL+NaOH+NaCl. (B) The sample's elemental composition predominantly includes Sodium (15.39 %), Oxygen (36.87 %), Carbon (47.38 %), and Chlorine (1.35 %). (C) C 1s high-resolution spectra of post-hydrothermal samples. Deconvolution of high-resolution spectra reveals the presence of sp³ hybridized carbon phase (47.8 %, 284.5 eV) in the post-hydrothermal samples, along with carbon in the functional groups such as -COOH, -C-O. (D) O 1s high-resolution spectra. The deconvolution of the O 1s spectra reveals the presence of absorbed Oxygen (531.1 eV) in the post-hydrothermal samples. (E) Na 1s high-resolution spectra. (F) Cl 2p high-resolution spectra.

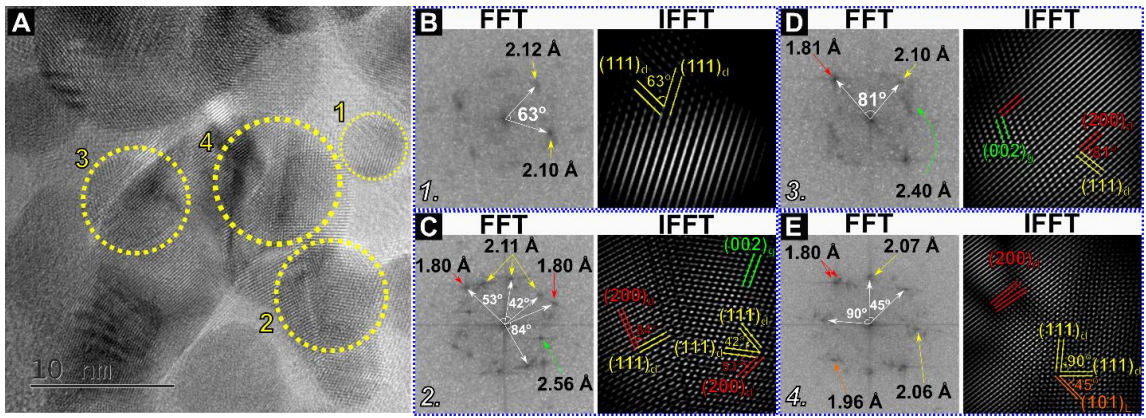

**Figure S26: HRTEM of the post-hydrothermal samples from Cellulose (CL)+NaOH+NaCl.** (A) determined the nanodiamond formation. (B) Particle 1 exhibits an atomic arrangement that forms an interplanar d-spacing of 2.10-2.12 Å and an interplanar angle of 63° (¡ 70°), indicating the formation of nanodiamonds with a stacking fault. (C) Particle 2 exhibits atomic arrangement, composing interplanar spacings of 2.11 Å and 1.80 Å corresponding to nanodiamond's (1 1 1) and (2 0 0) plane, while the lattice d- spacing of 2.56 Å indicates the constitution of graphitic carbon on nanodiamond particles. Particle 2 also showcases the disruption in atomic arrangement, causing a stacking fault in the nanodiamond. (D) The lattice d-spacing in particle 3 indicated the formation of a nanodiamond core with a graphitic carbon layer, where the nanodiamond exhibits a stacking fault due to the extension of the interplanar angle between (1 1 1) and (2 0 0) planes. (E) Particle 4 indicated the possible formation of lonsdaleite (hexagonal diamond) phases, where the lattice d-spacing of 2.06-2.07 Å composes an interplanar angle of 90°, corresponding to the atomic arrangement occurring in lonsdaleite diamonds. Further, the occurrence of lattice d- spacing of 1.96 Å corresponds to the lonsdaleite (1 0 1) plane. Overall, the nanodiamonds formed in the hydrothermal treatment of cellulose constitute a stacking fault and the formation of graphitic carbon at the surface. The lattice d-spacing of 2.08 − 2.12 Å and 1.80 − 1.81 Å, occurring at the angle of $\bar{5}5°$, correspond to nanodiamonds' (1 1 1) and (2 0 0) planes. Particle 2 constitutes a twinned nanodiamond crystal and graphitic outer layer, which depicts a contact twin caused by the rotation of the (1 1 1) plane of cubic diamond crystals. Particle 3 also exhibits the twinned nanodiamonds, where the reflection of (1 1 1) and (2 0 0) planes occur at 81°.